\documentclass[11pt]{article}

\usepackage{graphicx}
\begin{document}
\oddsidemargin .03in
\evensidemargin 0 true pt
\topmargin -.4in

%Abbreviations %
%***************%

\def\ra{{\rightarrow}}
\def\a{{\alpha}}
\def\b{{\beta}}
\def\l{{\lambda}}
\def\eps{{\epsilon}}
\def\T{{\Theta}}
\def\t{{\theta}}
\def\co{{\cal O}}
\def\car{{\cal R}}
\def\caf{{\cal F}}
\def\cs{{\Theta_S}}
\def\pr{{\partial}}
\def\tri{{\triangle}}
\def\na{{\nabla }}
\def\S{{\Sigma}}
\def\s{{\sigma}}
\def\sp{\vspace{.15in}}
\def\hs{\hspace{.25in}}

\newcommand{\be}{\begin{equation}} \newcommand{\ee}{\end{equation}}
\newcommand{\bea}{\begin{eqnarray}}\newcommand{\eea}
{\end{eqnarray}}

%********************************************************************%

\begin{titlepage}
\topmargin= -.2in
\textheight 9.5in

%\begin{flushright}
%{hep-th/0501}
%\end{flushright}
\begin{center}
\baselineskip= 18 truept

\vspace{.3in}
%\centerline{\Large\bf ${\mathbf AdS_5}$/CFT duality, Torsion and deformed $D_3$-brane %geometries:}
%\vspace{.01in}
%\centerline{\Large\bf Emergent gravity underlying a ${\mathbf U(1)}$ symplectic gauge theory}
%\vspace{.6in}
%\centerline{\Large\bf Stringy RN de Sitter black hole in 4D and Quintessence:} 
%\centerline{\Large\bf ${\mathbf{(3{\bar 3})}}$-brane Universe pair creation at a Big Bang}

\centerline{\Large\bf Quintessence and effective RN de Sitter brane geometries} 
%\centerline{\Large\bf Reissner-Nordstrom black hole on a vacuum created pair of ${\mathbf{(3{\bar 3})}}$-brane}

\vspace{.6in}
\noindent
{{\bf K. Priyabrat Pandey}, {\bf Abhishek K. Singh}, {\bf Sunita Singh},\\ {\bf Richa Kapoor} {\bf and} {{\bf Supriya Kar}\footnote{skkar@physics.du.ac.in }}}

\vspace{.2in}

\noindent
%{\large {${}^1$}The Abdus Salam International Centre for Theoretical Physics\\
%Strada Costiera 11, 34014 Trieste, Italy}

%\vspace{.15in}
%\centerline{\large and}
%\vspace{.15in}

\noindent
{{\Large Department of Physics \& Astrophysics}\\
{\Large University of Delhi, New Delhi 110 007, India}}

\vspace{.2in}

{\today}
\thispagestyle{empty}

\vspace{.6in}
\begin{abstract}
We revisit an effective space-time torsion curvature in a second order formalism, underlying the non-linear $U(1)$ gauge dynamics, of a two form on a $D_4$-brane in type IIA superstring theory. The formalism incorporates the significance of a global NS two form into the theory via its perturbative coupling to a dynamical two form. In particular, we explore the non-linear gauge dynamics on a $D_4$-brane in presence of a non-trivial background metric. The fact that the global modes of a NS two form in an open string theory sources the background metric on a $D_4$-brane may hint at the existence of an anti $D_4$-brane in the formalism. An effective de Sitter universe is shown to emerge on a vacuum created pair of $(D{\bar D})_3$-brane by a  local two form at the past horizon with a Big Bang. We obtain a number of $4D$ de Sitter quantum black holes, including a Reissner-Nordstrom (RN-) vacuum, with and without a propagating torsion. The quantum black holes are shown to be free from curvature singularity at $r\rightarrow 0$. In a low energy limit, the nonperturbative correction sourced by a torison may seen to be insignificant. The quantum black hole undergoes an expansion in the limit and presumably identifies with the Einstein vacuum. Interestingly our analysis reveals a plausible quintessence (axion) on an anti $D_3$-brane which may source the dark energy in a $D_3$-brane universe. Arguably a brane universe moves away from its anti-brane due to the conjectured repulsive gravity underlying the quintessence. It leads to a growth in extra fifth dimension between a brane and an anti-brane which may provide a clue behind an accelerating universe observed in cosmology.

\baselineskip=14 truept

\vspace{.12in}

\vspace{1in}
%\keywords {Near horizon D-brane, de Sitter tunneling, Emergent gravity, AdS brane, String theory, Torsion geometry.}
%\arxivnumber{1303.4344}
\noindent
%{\sc Published in:} {\bf JHEP 05 (2013) 033}

\noindent
%{\sc ArXiv:} {\tt 13034344} {[hep-th]}
\end{abstract}
\end{center}

\vspace{.2in}

\baselineskip= 16 truept

\vspace{1in}

%\noindent PACS: 11.25.-w; 11.15.-q

%\noindent Keywords: String theory, D-branes, Noncommutative space-time and Gravitation with Torsion

%\thispagestyle{empty}%
\end{titlepage}

\baselineskip= 18 truept

%%%%%%%%%%%%%%%%%%%%%%
\section{Introduction}
%%%%%%%%%%%%%%%%%%%%%% 
Dirichlet $p$-branes ($D_p$-branes) are established as the non-perturbative objects due to their Ramond-Ramond (RR) charges in type IIA or IIB superstring theory in ten dimensions \cite{polchinski}. The non-linear brane world-volume gauge dynamics is approximated by the Dirac-Born-Infeld (DBI) action. Since closed strings are tangential to a $D_p$-brane, Einstein gravity is believed to decouple from a $D_p$-brane world-volume \cite{seiberg-witten}. Nevertheless, a non-linear $U(1)$ gauge field is known to describe an effective open string metric on a $D_p$-brane and leads to a number of near horizon geometries  \cite{gibbons}-\cite{nicolini2}. In fact a constant Neveu-Schwarz (NS) two form, background in an open string theory, in combination with an electromagnetic field forms a gauge invariant field strength which governs a non-linear $U(1)$ charge on a $D_p$-brane. The effective geometries on a $D_p$-brane presumably provoke thought to investigate for an underlying effective (space-time) curvature theory in a second order formalism unparalleled to Einstein gravity.

\sp
\noindent
In the context it has been conjectured that a dynamical (NS) two form in an open string theory in principle may lead to a generalized non-commutative geometry on a $D_p$-brane. However the perspective of a dynamical two form on a $D_p$-brane has not been addressed due to some of the technical difficulties. In the recent past, we have explored a plausible effective gravity scenario, underlying a two form dynamics, on a $D_4$-brane \cite{spsk-JHEP}-\cite{spsk-NPB2}. Primarily the Poincare duality has played a significant role to identify a gauge theoretic torsion $H_3$ underlying the $U(1)$ gauge symmetry on a $D_4$-brane \cite{kpss}. Generically a dynamical higher form on a higher ($p>3$) 
dimensional $D_p$-brane may seen to re-define an open string effective metric. However in our analysis we have focused on a dynamical two form in the $U(1)$ gauge theory on a $D_4$-brane to construct a geometric torsion ${\cal H}_3$ in a five dimensional world-volume underlying a non-perturbative curvature. In fact a perturbative global NS two form coupling, to a local two form via its gauge curvature, has been shown to modify the covariant derivative appropriately in an underlying gauge theory. Alternately, the world-volume of a $D_4$-brane may be approximated by the dynamics of a geometric torsion in an effective space-time curvature conceptualized in a second order formalism. The $U(1)$ gauge invariance(s), underlying an effective curvature in a non-perturbative formalism and a non-linear gauge curvature in a perturbative theory, are checked to be restored independently. The two form $U(1)$ gauge invariance of the effective curvature scalar ensures a metric fluctuation and hence describes space-time in the formalism.

\sp
\noindent
On the other hand a torsion is known to play a significant role in the folk lore of string theories \cite{candelasHS,freed}. 
The torsion geometries leading to various quantum black holes in five dimensions have been obtained \cite{spsk-JHEP}. It is argued that the global modes of a NS two form in the string background may play an important role to describe a constant vacuum energy density in an effective curvature formalism. Thus a propagating two form together with the NS non-propagating modes are sensibly addressed in an effective curvature scalar 
${\tilde{\cal K}}$ underlying a geometric torsion. It was shown that in a gauge choice for a non-propagating torsion, the effective space-time on a $S^1$ may be identified with the Riemannian geometry \cite{spsk-PRD}. In fact the emergent brane geometries are analyzed for a torsion potential for its dependence on radial coordinate to confirm their stringy feature. In a low energy limit the effective torsion potential may be ignored. The quantum brane geometries in the limit presumably identify with some of the Einstein vacua.

\sp
\noindent
Interestingly a space-time (quantum) fluctuation on a $D_4$-brane in an effective curvature theory may alternately be viewed through a pair creation of a $D_3$-brane and an anti $D_3$-brane (${\bar D}_3$-brane) at a vacuum in a non-linear $U(1)$ gauge theory. For instance, we consider a dynamical two form in a $U(1)$ gauge theory on a $D_4$-brane in presence of a non-trivial background metric sourced by the global modes of NS two form in the open string bulk. The background metric for a $D_4$-brane gauge dynamics presumably describes a hidden ${\bar D}_4$-brane in the formalism. A priori, the $D_4$-brane vacuum may seen to be governed by an effective black hole essentially sourced by the global modes of a NS two form \cite{gibbons}-\cite{nicolini2}. Then the two form in the gauge theory may seen to vacuum create a pair of $(D{\bar D})_3$-brane at the past horizon of a background pure de Sitter. The pair creation by a two form in the gauge theory on a $D_4$-brane is analogous to a charged particle and an anti-particle pair production by a photon at the event horizon of a background black hole \cite{hawking}. The mechanism incorporates the quantum gauge theoretic effects into a classical black hole and enables one to address a semi-classical black hole and Hawking radiations. In our effective curvature formalism, the dynamical space-time has been argued to be created at a Big Bang singularity which turns out to describe a past horizon \cite{spsk-JHEP}. In other words, a dynamical two form in a $U(1)$ gauge theoretic vacuum on a $D_4$-brane creates a pair of $(D{\bar D})_3$-brane and they move away from the black hole event horizon with opposite momentum. The emergent space-time on a pair of brane/anti-brane is essentially sourced by an intrinsic non-linearity in a dynamical two form. An axion, Poincare dual to a torsion, on a ${\bar D}_3$-brane may seen to describe a quintessence underlying a varying vacuum energy density observed by CMB in cosmology. For instance, see references \cite{chen-jing-quintessence}-\cite{huan-jun-quintessence} for some of the recent developments with the idea of quintessence in literature. Qualitative analysis leading to a quintessence in the formalism may seen to be sourced by the hidden fifth dimension between a pair of vacuum created $(D{\bar D})_3$-brane.

\sp
\noindent
Furthermore the coupling of a global mode of NS two form to a gauge theoretic torsion ensures an emergent dynamical space-time via a metric fluctuation. Thus an effective geometry on a $D_4$-brane may alternately be viewed via a vacuum created pair of $(D{\bar D})_3$-brane Universe separated by an extra fifth dimension transverse to their four dimensional world-volumes. Generically a pair of lower dimensional brane/anti-brane space-time, from a higher dimensional $D$-brane vacuum, breaks supersymmetry and leads to a non-BPS D-brane configuration in an effective curvature formalism. Nevertheless from the perspective of a perturbative gauge theory, it may be revealing to analyze an appropriate Higgs mechanism underlying a spontaneous symmetry breaking. The broken symmetry  phase associated with the massive modes may describe a vacuum created pair of $(D{\bar D})_3$-brane.

\sp
\noindent
In the paper we obtain effective de Sitter brane geometries by geometric engineering of the ansatz for the global modes of a two form and a local two form. In particular, we consider two different plausible $4D$ scenario underlying a torsion free Universe and a new Universe in presence of a dynamical geometric torsion. The source potential in the de Sitter like black holes on a vacuum created $D_3$-brane and anti brane hint at a hidden fifth dimension in the formalism.

\sp
\noindent
On the one hand a torsion free universe is a special case underlying a gauge choice in the effective torsion curvature formalism. In the case, we obtain a Schwarzschild de Sitter and a topological de Sitter like black holes in $4D$ with a hint for an extra dimension. The effective de Sitter vacua are further analyzed for their angular velocities, event horizon and curvature singularities if any. We redefine the background, global NS two form, charge by absorbing the torsion charge and obtain a typical $4D$ Schwarzschild like black hole in the effective curvature formalism. For instance, we refer to references \cite{bousso-hawking}-\cite{lust} for an extensive review on various recent developments in de Sitter geometries and inflationary string cosmology.

\sp
\noindent
On the other a nontrivial torsion in the formalism is shown to modify the background de Sitter vacuum. The causal geometric patches in the brane geometry may appropriately be rearranged to yield an effective Reissner-Nordstorm (RN-) like de Sitter in presence of a perturbative correction underlying a gauge theoretic torsion charge coupled to a flat space-time. A priori the RN like de Sitter shall be seen to be sourced by three independent potentials underlying three non-linear charges in the formalism. They are: (i) the global modes of a NS two form, (ii) an axion (Poincare dual to a gauge theoretic $H_3$) and (iii) a geometric torsion. It is needless to mention that a non-linear charge sourced by a geometric torsion reassures the coupling of a global NS two form with a local two form in the formalism. The charge underlying a geometric torsion associates a stringy feature with the effective RN-like de Sitter on a vacuum created pair of $(D{\bar D})_3$-brane. The torsion charge is renormalized at the past horizon to describe a topological de Sitter like black hole on a brane world.

\sp
\noindent
We plan the paper as follows. We begin with a moderate introduction in section 1. It is followed by an out-line on a geometric torsion underlying a second order curvature formalism in section 2. An effective geometry scenario on a vacuum created pair of $(D{\bar D})_3$-brane at a Big Bang by a two form, in presence of a background metric, in a $U(1)$ gauge theory on a $D_4$-brane is presented in section 3. An axion on ${\bar D}_3$-brane, possessing its origin in a torsion, is qualitatively analyzed for its dual role leading to: (i) a quintessence vacuum energy density in section 3.1 and (ii) an D-instanton (non-perturbative) correction to the perturbative vacuum in section 3.2. We work out the effective de Sitter torsion free geometries on a brane in section 4, which are shown to describe a Schwarzschild de Sitter and a topological de Sitter in $4D$ with a hidden fifth dimension. In section 5 we work out new quantum de Sitter geometries underlying a propagating torsion. A stringy RN de Sitter black hole and a renormalization of a torison charge leading to topological de Sitter are obtained under the new geometries. We conclude in section 6 by summarizing the results with a plausible remark on a quintessence (axion) in the formalism.

%%%%%%%%%%%%%%%%%%%%%%%
\section{Preliminaries}
%%%%%%%%%%%%%%%%%%%%%%%
%%%%%%%%%%%%%%%%%%%%%%%%%%%%%%%%%%%%%%%%%%%%%%%%%%%%%%%%%%%%%%%%%%%%%%%
\subsection{Geometric torsion ${\mathbf{\cal H}_3}$ in ${\mathbf{5D}}$}
%%%%%%%%%%%%%%%%%%%%%%%%%%%%%%%%%%%%%%%%%%%%%%%%%%%%%%%%%%%%%%%%%%%%%%%
A $D_4$-brane is a dynamical object in ten dimensional type IIA superstring theory. Its five dimensional world-volume dynamics may be approximated by a Dirac-Born-Infeld (DBI) action which describes a non-linear one form ${\cal A}_{\mu}$ in a $U(1)$ gauge theory. For a constant background metric $g_{\mu\nu}$ the ${\cal A}_{\mu}$ dynamics, leading to a non-linear field strength ${\bar{F}}_{\mu\nu}=(2\pi\alpha')F_{\mu\nu}
= \left ({\bar F}_{\mu\nu}^{\rm linear} + B_{\mu\nu}^{\rm global}\right )$, is given by
\be
S= -{1\over{4C_1^2}}\int d^5x\ {\sqrt{-\det \left ( g +  {\bar{F}}\right )_{\mu\nu}}} \ ,\label{gauge-1}
\ee 
where $C_1^2=(4\pi^2g_s){\alpha'}^{5/2}$. Alternately the Poincare dual to the field strength on a $D_4$-brane underlie a dynamical two form $B_{\mu\nu}$, and is known to incorporate a gauge theoretic torison $H_3=dB_2$. A two form gauge dynamics is given by
\be
S=- {1\over{12C_2^2}}\int d^5x\ {\sqrt{-g}}\ H_{\mu\nu\lambda}H^{\mu\nu\lambda}\ ,\qquad {\rm where}\quad C_2^2=(8\pi^3g_s){\alpha'}^{3/2}\ .\label{gauge-2}
\ee
Interestingly the significance of the global modes of NS two form in addition to the local modes of a two form have been addressed in an effective curvature formalism on a $D_4$-brane \cite{spsk-JHEP,spsk-PRD}. A priori, a gauge theoretic torsion $H_3$ has been identified with a connection in the formalism leading to a modified covariant derivative ${\cal D}_{\mu}$ on a brane. The gauge connections are appropriately coupled to a two form and has been shown to define a geometric torsion ${\cal H}_3$ which takes the usual form:
\be
{\cal H}_{\mu\nu\lambda}= {\cal D}_{\mu}B_{\nu\lambda}\ +\;  {\rm cyclic\ in}\; (\mu, \nu, \lambda )\ ,\label{gauge-3}
\ee  
\be
{\rm where}\qquad {\cal D}_{\lambda}B_{\mu\nu}=\nabla_{\lambda}B_{\mu\nu} + {1\over2}H_{{\lambda\mu}}^{\rho}B_{\rho\nu} - {1\over2} H_{{\lambda\nu}}^{\rho}B_{\rho\mu}\ .\label{gauge-4}
\ee
An iterative incorporation of $B_2$-corrections, to all orders, in the covariant derivative leads to an exact derivative in a perturbative gauge theory. It may seen to define a non-perturbative covariant derivative in a second order formalism underlying a geometric realization. Then the geometric torsion may be given by
\bea
{\cal H}_{\mu\nu\lambda}&=&\nabla_{\lambda}B_{\mu\nu} + {1\over2}{{{\cal H}}_{\lambda\mu}}^{\rho}B_{\rho\nu} - {1\over2}{{\cal H}_{\lambda\nu}}^{\rho}B_{\rho\mu}\ + {\rm cyclic\ in}\; (\mu, \nu, \lambda)\ ,\nonumber\\
&=&H_{\mu\nu\lambda} + 3{{\cal H}_{[\mu\nu}}^{\alpha}{B^{\beta}}_{\lambda ]}\ g_{\alpha\beta}\nonumber\\
&=&H_{\mu\nu\lambda} + \left ( H_{\mu\nu\alpha}{B^{\alpha}}_{\lambda} + \rm{cyclic\; in\;} \mu,\nu,\lambda \right )\ +\ H_{\mu\nu\beta} {B^{\beta}}_{\alpha} {B^{\alpha}}_{\lambda}\ +\ \dots \; .\qquad\qquad\qquad {}\label{gauge-5}
\eea
The $U(1)$ gauge invariance of a geometric torsion under a two form transformation shall explicitly be realized in an effective space-time curvature formalism. Nevertheless in the modified $U(1)$ gauge theory, a global two form acts as a perturbation parameter and is distinctly placed than the local two form. Thus the gauge invariance $\delta_{\rm nz}{\cal H}_3=0$ may be retained by the non-zero modes of the two form as: $\delta_{\rm nz} H_3=0$ and 
$\delta_{\rm nz}B^{\rm global}=0$.

%%%%%%%%%%%%%%%%%%%%%%%%%%%%%%%%%%%%%%%%%%%%%%%%%%%%%%%%%%%%
\subsection{Space-time curvature on a ${\mathbf{D_4}}$-brane}
%%%%%%%%%%%%%%%%%%%%%%%%%%%%%%%%%%%%%%%%%%%%%%%%%%%%%%%%%%%%
Interestingly, a $B_2$-field dynamics in a first order formalism may be viewed as a torsion dynamics on an effective $D_4$-brane in a second order formalism \cite{spsk} leading to a fourth order generalized tensor:
\be
4{{\tilde{\cal K}}_{\mu\nu\lambda}{}}^{\rho}= 2\partial_{\mu}{{\cal H}_{\nu\lambda}}^{\rho} -2\partial_{\nu} {{\cal H}_{\mu\lambda}}^{\rho} + {{\cal H}_{\mu\lambda}}^{\sigma}{{\cal H}_{\nu\sigma}}^{\rho}-{{\cal H}_{\nu\lambda}}^{\sigma}{{\cal H}_{\mu\sigma}}^{\rho}.\label{gauge-6}
\ee
The generalized tensor is antisymmetric under an exchange of indices within a pair and is not symmetric under an exchange of its first pair of indices with the second. Hence, it differs from the Riemannian tensor $R_{\mu\nu\lambda\rho}$. However for a non-propagating torsion, 
${\tilde {\cal K}}_{\mu\nu\lambda\rho}\rightarrow R_{\mu\nu\lambda\rho}$. The second order curvature tensor and the irreducible curvature scalar are worked out to yield: 
\bea
4{\tilde{\cal K}}_{\mu\nu}&=& -\left (2\partial_{\lambda}{{\cal H}_{\mu\nu}}^{\lambda} +
{{\cal H}_{\mu\rho}}^{\lambda}{{\cal H}_{\lambda\nu}}^{\rho}\right )
\nonumber\\
{\rm and}\qquad {\tilde{\cal K}}&=& -{1\over{4}}{\cal H}_{\mu\nu\lambda}{\cal H}^{\mu\nu\lambda}
\ .\label{gauge-7}
\eea
As a result, the effective curvature constructed in a non-perturbative framework may be viewed as a generalized curvature tensor.
In a second order formalism an effective ${\cal H}_3$ dynamics on a $D_4$-brane may be approximated by
\be
S_{\rm D_4}^{\rm eff}= {1\over{3C_2^2}}\int d^5x {\sqrt{-\det {\tilde G}_{\mu\nu}}}\;\;\ {\tilde{\cal K}}\ .\label{gauge-6}
\ee
Generically the open string metric may arguably be re-expressed in terms of the Poincare dual field strength. It takes a form: 
\be
{\tilde G}_{\mu\nu}=\left ( {\tilde G}^z_{\mu\nu} + C\ {\bar{\cal H}}_{\mu\lambda\rho}{{\cal H}^{\lambda\rho}}_{\nu}\right )\ ,\label{gauge-7}
\ee
where ${\tilde G}^z_{\mu\nu}=\left ( g_{\mu\nu} - B_{\mu\lambda}{B^{\lambda}{}}_{\nu}\right )$ and $B_{\mu\nu}$ signify the global or zero modes of two form which are indeed coupled to an electromagnetic field $F_{\mu\nu}^{\rm linear}$. Under a two form gauge transformation, $\delta B_{\mu\nu}= \left (\partial_{\mu}\epsilon_{\nu} - \partial_{\nu}\epsilon_{\mu}\right )$, the $U(1)$ gauge invariance underlying the curvature scalar ${\tilde{\cal K}}$ is worked out explicitly to yield: 
\bea
\delta {\tilde{\cal K}} &=& {3\over2}\delta\left({\cal D}_{[\mu}B_{\nu\lambda ]}\right)\ {{\cal H}^{\mu\lambda\nu}}\nonumber\\
&=& {1\over2} \delta\left(H_{\mu\nu\lambda} + {H_{\mu\nu}}^{\rho} B_{\rho\lambda}+ {H_{\nu\lambda}}^{\rho} B_{\rho\mu}+ {H_{\lambda\mu}}^{\rho} B_{\rho\nu}\right)\ {{\cal H}^{\mu\lambda\nu}}\nonumber\\
&=&{3\over2} {H_{\mu\nu}}^{\rho}\ \delta {B_{\rho\lambda}}\ {{\cal H}^{\mu\lambda\nu}}\nonumber\\
&=&{3\over2}\left({H_{\mu\nu}}^{\lambda}\ {{\cal H}^{\mu\nu\rho}} - {H_{\mu\nu}}^{\rho}\ 
{{\cal H}^{\mu\nu\lambda}}\right)\partial_{\rho}\epsilon_{\lambda}\ .\label{gauge-8}
\eea
The $U(1)$ gauge invariance, $i.e.\ \delta {\tilde{\cal K}}=0$, may be achieved with a notion of a metric (fluctuation) sourced by the local degrees in a geometric torsion. A priori, the gauge invariance of the curvature scalar ${\tilde{\cal K}}$ seems to incorporate a condition on a two form:
\be
C\ (2\pi\alpha'){H_{\alpha\beta}}^{\mu}\ {{\cal H}^{\alpha\beta\nu}} = f^{\mu\nu}_q\ .\label{gauge-9}
\ee
Nevertheless the perturbation gauge theory re-assures an infinitesimally small global (two form) mode coupling in a geometric torison (\ref{gauge-5}) and hence the condition may seen to be ignored. Then the emerging notion of a metric underlying a propagating geometric torsion in the non-perturbative curvature formalism is given by
\be
f_{\mu\nu}^q\ =\ C\ {\cal H}_{\mu\alpha\beta} {{\bar{\cal H}}^{\alpha\beta}}_{\nu}\ .\label{gauge-10}
\ee
The symmetricity under an interchange $\mu\leftrightarrow \nu$ in the r.h.s. ensures that the naive condition on a two form becomes insignificant. Interestingly, the emergent metric incorporates the notion of space-time in the curvature scalar ${\tilde{\cal K}}$ and hence justifies the open string effective metric arguably obtained in eq.(\ref{gauge-7}) on a $D_4$-brane.

%%%%%%%%%%%%%%%%%%%%%%%%%%%%%%%%%%%%%%%%%%%%%%%%%%%%%%%%%%%%%%%
\section{Vacuum created pair of $\mathbf{(D{\bar D})_3}$-brane}
%%%%%%%%%%%%%%%%%%%%%%%%%%%%%%%%%%%%%%%%%%%%%%%%%%%%%%%%%%%%%%%

It is remarkable to note that an effective space-time is vacuum created on a pair of $(D{\bar D})_3$-brane by a dynamical two form in a $U(1)$ gauge theory on a $D_4$-brane. The vacuum refers to a background black hole on a $D_4$-brane which is known to be sourced by a global mode of a NS two form. Hence a   geometric torsion, underlying a second order curvature formalism, is associated with a vacuum created pair of $(D{\bar D})_3$-brane Universe. In fact the dynamical space-time without a torison in $4D$ has been argued to began at a Big Bang singularity, defined at a cosmological horizon, in a $5D$ braneworld with a torsion. It is noteworthy to remark that the space-time curvature, sourced by a geometric torsion in $5D$, naturally incorporates an intrinsic angular momentum on a vacuum created pair of $(D{\bar D})_3$-brane. The world-volumes of a brane and an anti-brane are separated by an extra fifth dimension transverse to each other. It may imply that an observer in a $D_3$-brane universe is inaccessible to an anti $D_3$-brane Universe or vice-versa. Nevertheless the invisible affect of an anti $D_3$-brane space-time is indeed compelling on a 3-brane Universe due to the hidden fifth dimension. Analysis may reveal that a brane-Universe geometry seems to be influenced by a potential candidate for dark energy.

%%%%%%%%%%%%%%%%%%%%%%%%%
\subsection{Quintessence}
%%%%%%%%%%%%%%%%%%%%%%%%%
The CMB observations in cosmology assures that the vacuum energy density is not a constant. A time variation of cosmological vacuum energy density in the early epoch leading to a de Sitter geometry has been argued in the inflationary models. In the context, a quintessence is known to describe a time varying, spatially homogenous component of cosmic density with a negative pressure and a positive energy density. In the recent past, a slowly rolling time dependent scalar field is believed to describe the quintessence presumably due to its simplest form as a tensor field which couples minimally to Einstein gravity. The acceleration of universe is parameterized by the cosmological equation of state: $\omega=P/\rho$, where $P=-T^i_i<0$ and $\rho= T^0_0>0$ respectively defines the pressure and energy density sourced by a scalar field. In the recent years the observational cosmology primarily focuses on the measurement of $\omega$ sourced by a quintessence. It takes a bounded value: $-1<\omega<-1/3$.

\sp
\noindent
Quintessence energy density, underlying a geometric torsion, in the formalism may seen to be sourced by a dynamical two form. For instance, a non-trivial curvature ${\cal K}$ may seen to be sourced by a dynamical mass-less axion on an anti $D_3$-brane. The presence of an extra fifth dimension transverse to the anti-brane affects a vacuum created $D_3$-brane universe which is described by a non-linear two form ${\cal F}_2$. The inaccessible scalar dynamics to an observer in an effective space-time on a $D_3$-brane provokes thought to identify the Poincare dual of torsion as a quintessence scalar field into the formalism. The idea of a quintessence energy density sourced by an axion field on an anti-brane is remarkable. A quintessence scalar field is believed to be a potential candidate to describe the dark energy in our universe. The fact that the dark energy permeats all the space and tends to enhance the rate of expansion of universe may be understood via an extra transverse dimension between a pair of vacuum created $D_3$-brane and ${\bar D}_3$-brane in the formalism. Hubble's law may be revisited to explain the receding of a brane from an anti-brane. The law states that the recessional speeds of a brane and anti-brane universes are proportional to the transverse distance between them. Since a vacuum created brane universe moves away from its an anti-brane in an opposite direction, it implies a repulsive effective gravity between them. The notion of repulsive gravity has been argued to be sourced by a quintessence scalar field in cosmology. It is thought provoking to believe that an axion in a hidden anti-brane is realized through a geometric torsion, which in turn is responsible for the repulsive gravity. In fact a repulsive gravity underlying a quintessence has been conjectured between a matter and anti-matter which may seen to be in conformity with the notion of a vacuum created pair of $(D{\bar D})_3$-brane by a two form on a $D_4$-brane. Interestingly the effective de Sitter black holes, obtained in a gauge choice for a nonpropagating torsion, may seen to be described with a varying positive curvature scalar $R=12/r^4$. It further strengthen the presence of quintessence axion in the formalism. It may imply that the brane/anti-brane universe(s) at its vacuum creation, $i.e.$ at the past horizon radius $b$ with a Big Bang, was described by a constant curvature $R=12/b^4>0$. At a Big Bang the quintessence kinetic energy is ignored and hence the equation of state reduces to $\omega=-1$, The result is in agreement with the value of $\omega$ known for a constant vacuum energy density.

%%%%%%%%%%%%%%%%%%%%%%%%%%%%%%%%%%%%%%%%%%%%%%
\subsection{$\mathbf{D}$-instanton correction}
%%%%%%%%%%%%%%%%%%%%%%%%%%%%%%%%%%%%%%%%%%%%%%
An axion in the Ramond-Ramond sector of type II superstring hints for a D-instanton in the formalism. Importantly a D-instanton is known to incorporate a non-perturbative correction in a perturbative string theory. In fact the universe was argued to begin at a past horizon with a Big Bang on a pair of 
$(D{\bar D})$-instanton at an early epoch in de Sitter underlying a geometric torsion formalism \cite{spsk-JHEP}. The hot de Sitter is believed to expand and the notion of a real time was incorporated subsequently in the geometry. The expansion of a pair of brane (or anti-brane) universe from its origin at a Big Bang was argued to go through a series of nucleation a pair of $D$-instanton followed by higher dimensional pairs: $D$-particle, $D$-string, $D$-membrane and process stops with a pair of $D_3$-brane in the formalism. It is due to the fact that a pair of $(D{\bar D})_3$-brane with a transverse fifth dimension completely fills the $D_4$-brane world-volume.

\sp
\noindent
In other words an instanton local degree on an anti $D_3$-brane influences the world-volume of a $D_3$-brane within a pair. Arguably the instantaneous affect of a ${\bar D}$-instanton on a $D_3$-brane is in agreement with the non-perturbative universe in $4D$. For $\kappa^2=(2\pi)^{5/2}g_s\alpha'$, the non-perturbative curvature dynamics on $S^1$ takes a form:
\be
S= {1\over{3\kappa^2}}\int d^4x {\sqrt{-\det G_{\mu\nu}}}\;\;\ \left ( {\cal K}\ -\ {3\over4} {\bar{\cal F}}_{\mu\nu}
{\cal F}^{\mu\nu} \right )\ ,\label{gauge-11}
\ee
\be
{\rm where}\qquad {\cal F}_{\mu\nu}=\ F_{\mu\nu}\ +\ {\cal H}_{\mu\nu}^{\lambda}{\cal A}_{\lambda}\ .\qquad\qquad\qquad\qquad\qquad\qquad\qquad
\qquad\qquad {}\label{gauge-12}
\ee
The curvature scalar ${\cal K}$ is essentially sourced by a dynamical two form potential in a first order formalism. Its field strength is appropriately modified to define a dynamical geometric torsion ${\cal H}_3$ in four dimensions underlying a second order space-time curvature formalism. The fact that a torsion is dual to an axion (scalar) on an effective $D_3$-brane, ensures one degree of freedom. In addition, ${\cal F}_{\mu\nu}$ describes a geometric one form field with two local degrees on an effective $D_3$-brane. A precise match among the (three) local degrees of torsion in ${\tilde{\cal K}}$ on $S^1$ with that in ${\cal K}$ and ${\cal F}_{\mu\nu}$ reassure the absence of a dynamical dilation field in the frame-work. The result is consistent with the fact that a two form on $S^1$ does not generate a dilation field. Alternately an emergent metric tensor, essentially sourced by a dynamical two form, does not formally evolve under a compactification of an underlying $U(1)$ gauge theory (\ref{gauge-6}) on $S^1$.

\sp
\noindent
In absence of a geometric torsion, eq(\ref{gauge-12}) reassures a gauge theoretic non-linear field strength on a $D_3$-brane. Intuitively, a geometric torsion leading to a spinning $D_3$-brane presumably enforces the notion of an anti $D_3$-brane spinning in an opposite direction to the brane. In a low energy limit, the spin slows down significantly and may be ignored to retain the BPS configuration, $i.e.$ a BPS brane and an anti BPS brane. In the limit the extra fifth dimension becomes large and hence a brane Universe and an anti-brane Universe may appear to be independent of each other. In the case, the quintessence field dynamics freezes on an anti brane and the brane Universe is precisely governed by two local degrees of the gauge theoretic non-linear one form.

\sp
\noindent
On the other hand, the energy-momentum tensor $T_{\mu\nu}$ is computed in a gauge choice: $${3\over4}{\bar{\cal F}}_{\mu\nu}{\cal F}^{\mu\nu}= {3\over{\pi\alpha'}} + {\cal K}\ .$$ 
Explicitly
\be
(2\pi\alpha')^2T_{\mu\nu}=\left ( G^z_{\mu\nu} + {\tilde C}\ {\bar{\cal F}}_{\mu\lambda}
{{\bar{\cal F}}^{\lambda}{}}_{\nu}+ C\ {\bar{\cal H}}_{\mu\lambda\rho}{{\cal H}^{\lambda\rho}{}}_{\nu}\right )\ .\label{gauge-13}
\ee
The gauge choice ensures that the $T_{\mu\nu}$ sources a nontrivial emergent geometry underlying a non-linear $U(1)$ gauge theory. Interestingly, the $T_{\mu\nu}$ sources an effective metric $G_{\mu\nu}$ in the frame-work. 
$T_{\mu\nu}$, with ($C={3\over4}$, ${\tilde C}={3\over2}$) and ($C=-{5\over4}$, ${\tilde C}=-{1\over2}$), respectively correspond to a $(+)$ve sign and a $(-)$ve sign in $G_{\mu\nu}$. A priori, they describe two inequivalent black holes on an effective $D_3$-brane. Two solutions for an emergent metric tensor is a choice keeping the generality in mind. Primarily, they dictate the quantum geometric corrections to a background metric.
Generically, the effective metric on a $D_p$-brane for $p\neq 4$ may be given by
\be
G_{\mu\nu}=\left ( G^z_{\mu\nu} \pm {\bar{\cal F}}_{\mu\lambda}{{\bar{\cal F}}^{\lambda}{}}_{\nu} \pm 
{\bar{\cal H}}_{\mu\lambda\rho}{{\cal H}^{\lambda\rho}{}}_{\nu} \right )\ .\label{gauge-14}
\ee
The generalized metric incorporate the local modes of a two form in addition to the global modes of a NS two form from the open string bulk. A torsion may seen to incorporate conserved quantities, $i.e.$ angular momentum and/or a charge, on a brane-world. Interestingly, the torsion brane-world geometries become significant in the Planck energy scale and may be identified with a primordial rotating and/or charged black holes and their higher dimensional generalizations.

%%%%%%%%%%%%%%%%%%%%%%%%%%%%%%%%%%%%%%%%%%%%%%%%%%%%%%%
\section{Quantum de Sitter brane in 4D without torsion}
%%%%%%%%%%%%%%%%%%%%%%%%%%%%%%%%%%%%%%%%%%%%%%%%%%%%%%%
%%%%%%%%%%%%%%%%%%%%%%%%%%%%
\subsection{Two form ansatz}
%%%%%%%%%%%%%%%%%%%%%%%%%%%%
A geometric torsion, underlying a $U(1)$ gauge theory, on a $D_4$-brane may alternately be viewed via a vacuum created pair of $(D{\bar D})_3$-brane. Within a pair the world volumes are distinctly separated by an extra fifth dimension. In other words, an irreducible scalar curvature ${\tilde{\cal K}}$ on $S^1$ generically reduces to ${\cal K}$ on a $D_3$-brane and a geometric ${\cal F}_2$ on an anti $D_3$-brane or vice-versa. The two form ansatz in $4D$ may be given by
\bea
B_{t\theta}&=& B_{r\theta} = {{b}\over{\sqrt{2\pi\alpha'}}}\ ,\nonumber\\
B_{\theta\phi}&=& {{p}\over{\sqrt{2\pi\alpha'}}} \sin^2\t\ , \nonumber\\
{\cal A}_t&=& {{Q_e}\over{r}}\nonumber\\
{\rm and}\qquad {\cal A}_{\phi}&=& - {{Q_m}\over{\sqrt{2\pi\alpha'}}} \cos\t\ ,\label{ansatz-1}
\eea
where $(b,p)>0$ are constants and $(Q_e,Q_m)$ denote the non-linear (electric, magnetic) charges. The gauge choice re-assures a vanishing torsion $H_3=0={\cal H}_3$ in presence of the $B_2$-fluctuations. The two form field strength is worked out for its non-vanishing components to yield:
\be
{\cal F}_{0r} = {{Q_e}\over{r^2}}\qquad {\rm and}\quad {\cal F}_{\t\phi}= {{Q_m}\over{2\pi\alpha'}}\sin\t\ .\label{ansatz1a}
\ee
The emergent quantum geometry, in a low energy limit, may describe a four dimensional Einstein vacuum. The open string effective metric sourced by a constant two form may seen to receive further geometric corrections due to the non-linear $U(1)$ charges underlying a generic ${\cal F}_2$ on a $D_3$-brane. Generically, the emergent metric may be expressed as:
\be
{G}_{\mu\nu}=\left ( g_{\mu\nu}\ \pm\ B_{\mu\lambda}g^{\lambda\rho}B_{\rho\nu}\ \pm\ {\bar{\cal F}}_{\mu\lambda}g^{\lambda\rho}{\bar {\cal F}}_{\rho\nu}\right )\ .\label{emetric}
\ee
For simplicity we set $(2\pi\alpha')=1$. Explicitly the line element(s), underlying the nontrivial brane geometries, are given by
\bea
ds^2=&-&\left (1 - {{b^2}\over{r^2}} \pm {{Q_e^2}\over{r^4}}\right ) dt^2 
+ \left ( 1 + {{b^2}\over{r^2}} \pm {{Q_e^2}\over{r^4}}\right ) dr^2 + 
{{2b^2}\over{r^2}} dt dr - {{2bp\sin^2\theta}\over{r^2}}\left ( dt + dr \right )d\phi \nonumber\\
&+&\left ( 1 + {{p^2\sin^2 \theta \mp Q_m ^2 }\over{r^4}}\right )  r^2 d\Omega^2\label{bh-1}
\eea 
and 
\bea
ds^2=&-&\left (1 + {{b^2}\over{r^2}} \pm {{Q_e^2}\over{r^4}}\right ) dt^2 
+ \left ( 1 - {{b^2}\over{r^2}} \pm {{Q_e^2}\over{r^4}}\right ) dr^2 - 
{{2b^2}\over{r^2}} dt dr + {{2bp\sin^2\theta}\over{r^2}}\left ( dt + dr \right )d\phi \nonumber\\
&+&\left ( 1 - {{p^2\sin^2 \theta \pm Q_m ^2 }\over{r^4}}\right )  r^2 d\Omega^2\ .\label{bh-2}
\eea 

%%%%%%%%%%%%%%%%%%%%%%%%%%%%%%%%%%%%%%%%%%%%%%%%%%%
\subsection{Hot de Sitter brane geometries}
%%%%%%%%%%%%%%%%%%%%%%%%%%%%%%%%%%%%%%%%%%%%%%%%%%%
We perform an appropriate Weyl scaling of the emergent metric to access de Sitter vacua without any change in causal properties. It is given by
\be
{\tilde G}_{\mu\nu}= \lambda(r) G_{\mu\nu}\ ,\qquad\qquad {}\label{dS4D-1}
\ee
where $\lambda = b^2/r^2$. For an allowed range of parameters $b>r>Q_e/b>0$, the conformal factor $\lambda >1$. The upper bounds on $r$ and $Q_e$
further enforce an ultra high energy scale in the transformed vacua. Though it implies a critical value for an electric field, it does not bound a magnetic field in the quantum vacua. Thus the magnetic field can be infinitely large and may seen to incorporate a rotation into the geometries.
The stringy geometries are worked out on a vacuum created pair of a $(D{\bar D})_3$-brane by a two form on a $D_4$-brane. They are
given by
\bea
ds^2&=&-\ \left (\pm 1+ {{r^2}\over{b^2}} \pm {{Q_e^2}\over{r^2 b^2}}\right ) dt^2 + 
\left ( \mp 1 + {{r^2}\over{b^2}} \pm {{Q_e^2}\over{r^2b^2}}\right ) dr^2\ \mp\ 2drdt \pm {{2p}\over{b}}\left ( dt + dr \right )\sin^2\t\ d\phi\nonumber\\
&&+\ {{r^4}\over{b^2}}\left (1 \mp {{p^2 \sin^2\theta + Q_m^2}\over{r^4}}\right)d\Omega^2\ .\label{dS4D-2}
\eea
We consider a $(-)$ve sign with a two form fluctuation in the emergent metric (\ref{emetric}). In the case, the stringy vacua may be given by
\bea
ds^2&=&-\ \left (- 1+ {{r^2}\over{b^2}} \pm {{Q_e^2}\over{r^2 b^2}}\right ) dt^2\ +\  
\left (  1 + {{r^2}\over{b^2}} \pm {{Q_e^2}\over{r^2b^2}}\right ) dr^2\ +\ 2drdt - {{2p}\over{b}}\left ( dt + dr \right )\sin^2\t\ d\phi\nonumber\\
&&\qquad\qquad\qquad\qquad\qquad\qquad\qquad\qquad\qquad\qquad\qquad\qquad\qquad\quad +\ \left ( \rho_+^2 \mp N_m^2\right )d\Omega^2\ .\label{dS4D-3}
\eea
Generically the modified radii of $S_2$ are: $\rho_{\pm}={1\over{b}}{\sqrt{r^4 \pm p^2\sin^2\t}}$.
In a brane regime $b>r>N_e$, the stringy geometries describe de Sitter black holes with euclidean signature on a $D_3$-brane. They are given by
\bea
ds^2&=&\left ( 1- {{r^2}\over{b^2}} \mp {{Q_e^2}\over{r^2 b^2}}\right ) dt^2\ +\ 
\left (  1 - {{r^2}\over{b^2}} \mp {{Q_e^2}\over{r^2b^2}}\right )^{-1} dr^2\ +\ 2drdt - {{2p}\over{b}}
\left ( dt + dr\right )\sin^2\t\ d\phi\nonumber\\
&&\qquad\qquad\qquad\qquad\qquad\qquad\qquad\qquad\qquad\qquad\qquad\qquad\quad\quad +\ \left ( \rho_+^2 \mp N_m^2\right )d\Omega^2\ .\label{dS4D-65}
\eea
Interestingly Weyl scaling of the metric may seen to describe an ultra high energy domain on a specified brane window. The scaling incorporates a change in signature of time in de Sitter geometries. A signature change ensures a temperature and presumably describes a hot de Sitter stringy vacuum \cite{spsk-JHEP}. The parameter `$b$' in the regime may be identified with a cosmological scale in the formalism. The difference, between an ultra-high and a high energy scale, incorporates a small positive vacuum energy density. It elevates the charged quantum black holes to the de Sitter stringy vacua non-perturbatively. A limit $\lambda\rightarrow 1$, underlying a Big Bang singularity, enforces $Q_e\rightarrow 0$ at the creation of the brane-Universe. In a global scenario, $i.e.$ on a pair of $(D{\bar D})_3$-brane, the quantum geometries in the regime with real spin angular momenta may reduce to yield:
\be
ds^2=-\left ( 1- {{r^2}\over{b^2}}\mp {{N_e^2}\over{r^2}}\right ) dt^2\ +\ \left (  1 - {{r^2}\over{b^2}} \mp {{N_e^2}\over{r^2}}\right )^{-1}dr^2
\ -\ {{2p}\over{b}} \sin^2\t\ d\phi dt\ +\ \left ( \rho_+^2 \mp N_m^2\right )d\Omega^2\ .\label{dS4D-5}
\ee 
The conserved electric and magnetic non-linear charges are, respectively, redefined as the point charges $N_e=Q_e/b$ and $N_m=Q_m/b$. Then, the quantum regime may be re-expressed as $b>r>N_e$.  A small $N_e$ or $N_m$ leading to a point charge is reassured by a cosmological scale $b>(Q_e, Q_m)$. In absence of electric and magnetic point charges, the geometries reduce to a de Sitter with a real time. In presence of charges ($N_e$ and $N_m$), they may seen to describe a Schwarzschild de Sitter and a topological de Sitter black hole in lorentzian singnature with a real angular momentum $\Omega^{\phi}$. A priori, an $S_2$-symmetric geometric patch associated with a magnetic charge appears to decouple from the quantum geometries. However in its presence, the quantum black holes may be re-expressed via a de Sitter sourced by a two form and a non-perturbative quantum correction sourced by a electric and a magnetic charge. Explicitly, they are given by
\be
ds^2=-\left ( 1- {{r^2}\over{b^2}}\right ) dt^2\ +\ \left (  1 - {{r^2}\over{b^2}}\right )^{-1} dr^2\ -\ {{2p}\over{b}} \sin^2\t\ d\phi dt\ +\ 
\rho_+^2 d\Omega^2\ \pm\ {{N_e^2}\over{r^2}}ds^2_{\rm flat}\ ,\label{dS4D-6}
\ee
where $ds^2_{\rm flat}=\left (dt^2 + dr^2 +r^2 d\Omega^2\right )$ and $Q_m=iQ_e$. In absence of $B_{\t\phi}$ fluctuation, $i.e.\ p=0$, the two form ansatz (\ref{ansatz-1}) primarily accounts for the causal patches of a pure de Sitter in $4D$ with a cosmological horizon at $r\rightarrow r_c=b$. Generically, the ansatz ($p\neq 0$) sources an angular velocity $\Omega^{\phi}$ which incorporates a spin angular momentum in the quantum de Sitter. 
The hot de Sitter vacua in eqs(\ref{dS4D-3}) and (\ref{dS4D-65}) may be analyzed for a Big Bang singularity in a limit $Q_e\rightarrow 0$. The effective de Sitter metric becomes singular there. Thus a pair of $(D{\bar D})_3$-brane was vacuum created with nontrivial torsion geometries at a cosmological horizon with a Big Bang. A priori the Riemannian curvature scalar is worked out for the brane/anti-brane together in a limit $Q_e\rightarrow 0$ to yield:
\be
R=\left ( {{34}\over{b^2}} -{{16}\over{r^2}} + {{2b^2}\over{r^4}}\right )\ .\label{curvature}
\ee
At the cosmological horizon $r\rightarrow b$ the curvature reduces to a small constant value $R=20/b^2$. 
It may imply that the brane universe began with a Big Bang. It was described by a small positive cosmological constant and hence lead to a de Sitter vacuum. Right after Big Bang the torsion in five dimensions played a significant role to source an one form on a $D_3$-brane. It is important to note that an electric non-linear charge, underlying an one form, couples to a flat metric in eq(\ref{dS4D-6}). It may be identified with a non-perturbative quantum correction presumably to Einstein vacuum in the formalism.

%%%%%%%%%%%%%%%%%%%%%%%%%%%%%%%%%%%%%%%%%%%%%%%%%%%%
\subsection{Schwarzschild and topological de Sitter}
%%%%%%%%%%%%%%%%%%%%%%%%%%%%%%%%%%%%%%%%%%%%%%%%%%%%
We analyze the effective de Sitter black holes in eq(\ref{dS4D-5}) in this section. They describe a Schwazschild de Sitter and a topological de Sitter geometries and are respectively given by 
\be
ds^2_{\rm QdS}=-\left ( 1- {{r^2}\over{b^2}} - {{N_e^2}\over{r^2}}\right ) dt^2 + \left (  1 - {{r^2}\over{b^2}} - {{N_e^2}\over{r^2}}\right )^{-1} dr^2
- {{2p}\over{b}}\sin^2\theta d\phi\ dt + \left ( \rho_+^2 - N_m^2\right )d\Omega^2\label{dS4D-30}
\ee
\be
{\rm and}\;\; ds^2_{\rm QdT}=-\left ( 1- {{r^2}\over{b^2}} + {{N_e^2}\over{r^2}}\right ) dt^2+\left (  1 - {{r^2}\over{b^2}} + {{N_e^2}\over{r^2}}\right )^{-1} dr^2 -{{2p}\over{b}}\sin^2\theta d\phi\ dt + \left ( \rho_+^2 + N_m^2\right )d\Omega^2.\label{dS4D-31}
\ee
The $4D$ causal de Sitter patches ensure an extra (fifth) dimension between a pair of $(D{\bar D})_3$-brane in the formalism. In particular, a 
Schwarzschild dS quantum black hole is characterized by two horizons at
\be
r_{\pm}={{r_c}\over{\sqrt{2}}}\left ( 1 \pm {\sqrt{1-{{4Q_e^2}\over{b^4}}}}\right )^{1/2}.\label{dS4D-32}
\ee
In the regime, the horizon radii reduces to yield:
\be
r_+\approx r_c\left (1 - {{N_e^2}\over{2r_c^2}}\right )\qquad {\rm and}\quad r_-\approx N_e\ .\label{dS4D-33}
\ee
The topological dS quantum black hole is characterized by an event horizon at
\bea
r_{e}&=&{{r_c}\over{\sqrt{2}}}\left ( 1 + {\sqrt{1+{{4Q_e^2}\over{b^4}}}}\right )^{1/2}.\nonumber\\
&\approx&r_c\left (1 + {{N_e^2}\over{2r_c^2}}\right )
\label{dS4D-35}
\eea
Thus the event horizons in a Schwarzschild dS and in a topological dS quantum black hole, respectively, be approximated by ($r_c-\epsilon$) and ($r_c+\epsilon$), where $\epsilon<1$. However, their radii in $S_2$ appears to shrink or expand in presence of a two form fluctuation and a magnetic charge. The event horizon is separated from a cosmological horizon by a time-like separation in a Schwarzschild dS quantum black hole. A real spin (angular momentum) turns out to be sourced by an electric charge in the de Sitter black holes. The angular velocities are computed at the event horizon(s) and are given by
\bea
&&\Omega^{\phi}_{\rm QdS}= {{-bp}\over{(b^2-N_e^2)^2 + (p^2 \sin^2\t -Q_m^2)}} \nonumber\\
{\rm and}&&\Omega^{\phi}_{\rm QdT}={{-bp}\over{(b^2+N_e^2)^2 + (p^2 \sin^2\t +Q_m^2)}}\ .\label{dS4D-5c}
\eea
It ensures a higher spin in a quantum Schwarzschild de Sitter than in a topological de Sitter black hole. A priori, the angular velocities appear to vary with the polar angle, which is sourced by a two form fluctuation $B_{\t\phi}$ in the formalism. At the creation of a pair of $(D{\bar D})_3$-brane, $i.e.$ at poles, the two form fluctuation vanishes and the angular velocities turn out to be maximum. The spins reduce to a minimum on the equator. The quantum de Sitter black holes presumably enforce a slicing in the polar angle, which may be a consequence of quantized space. A priori, the spinning de Sitter quantum black holes do not describe an ergo sphere though they are characterized by a conserved angular momentum $p$. However their radial coordinate is modified $r\rightarrow r_{\rm eff}$ in presence of the conserved charges ($p,Q_m$). They reflect the characteristics of a charged 
$4D$ black hole in string theory \cite{garfinkle}. In addition the effective radii in the quantum de Sitter black holes
in eqs(\ref{dS4D-30}) and (\ref{dS4D-31}) vary with a polar angle and hence deform a sphere to an ellipsoid. The radii increase from their minimum values at the poles to the maximum value on the equator. Formally, an effective radius on the equator may be identified with an ergo radius for a stringy Schwrazschild or topological black hole. It may be checked under an interchange $dt_e\leftrightarrow d\t$ and/or $dt_e\leftrightarrow d\phi$ with appropriate normalizations in de Sitter quantum black holes. In other words, a stringy de Sitter/topological black hole undergoes an expansion from poles to the equator.

\sp
\noindent
In a near (cosmological) horizon regime the $S_2$ radii may be approximated to the original radial coordinate $r$. In the limit the quantum black holes (\ref{dS4D-30}) and (\ref{dS4D-31}) may be approximated by their near horizon geometries. Incorporating a plausible flip in light cone in the near horizon de Sitter black holes, we obtain
\be
ds^2_{\rm QdS}\ \rightarrow\ -\ {{r^2}\over{N_e^2}}dt^2\ +\ {{N_e^2}\over{r^2}}dr^2\ -\ {{2p}\over{b}}\sin^2\theta\ d\phi dt\ +\ r^2 d\Omega^2\label{dS4D-34}
\ee
\be
{\rm and}\qquad\quad ds^2_{\rm QdT}\ \rightarrow\ -\ {{r^2}\over{N_m^2}} dt^2\ +\ {{N_m^2}\over{r^2}}dr^2\ -\ {{2p}\over{b}} \sin^2\theta\ d\phi dt\ +\ r^2 d\Omega^2\ .\qquad\qquad {}\label{dS4D-34b}
\ee
It may imply that a Schwrazschild and a topological de Sitter quantum black hole in their near (cosmological) horizon may be viewed through an asymptotic rotating AdS-brane defined, respectively, with the AdS radii $N_e$ and $N_m$. The correspondences between an asymptotic AdS quantum black holes and the near (cosmological) horizon Schwarzschild and topological black holes may enforce be a quantum tunneling: dS$\rightarrow$ AdS in the formalism. In fact, a tunneling has been argued from a Schwarzschild de Sitter quantum black hole to a Schwarzschild AdS via a topological de Sitter \cite{spsk-JHEP}.

\sp
\noindent
A priori an $S_2$-symmetric patch associated with a magnetic (non-linear) charge decouples from the stringy black holes. However we begin with the
dyonic charge quantum black holes and work out in a special case for ${p \sin\t =\pm Q_m}$. Formally with an euclidean time, they may be given by
\be
ds^2_{\rm QdS}=\left ( 1- {{r^2}\over{b^2}} - {{N_e^2}\over{r^2}}\right ) dt_e^2 + \left (  1 - {{r^2}\over{b^2}} - {{N_e^2}\over{r^2}}\right )^{-1} dr^2
\mp\ 2N_m \sin\theta\ d\phi dt\ +\ {{r^4}\over{b^2}} d\Omega^2\ ,\qquad\qquad\quad {}\label{dS4D-4a}
\ee
\be
ds^2_{\rm QdT}=\left ( 1- {{r^2}\over{b^2}} + {{N_e^2}\over{r^2}}\right ) dt_e^2 + \left (  1 - {{r^2}\over{b^2}} + {{N_e^2}\over{r^2}}\right )^{-1} dr^2
\mp\ 2N_m \sin\theta\ d\phi dt\ +\ \left ({{r^4}\over{b^2}} + 2N_m^2\right )d\Omega^2\ . \label{dS4D-4b}
\ee
Since $N_m^2d\Omega^2$ decouples from the rest of the de Sitter topological quantum black hole, both the geometries differ in their sign in the potential sourced by a small electric charge $N_e$ in the formalism. With a lorentzian signature, the angular velocity $\Omega^{\phi}$ is real and the quantum black holes are given by
\be
ds^2_{\rm QdS}=-\left ( 1- {{r^2}\over{b^2}} - {{N_e^2}\over{r^2}}\right ) dt^2 + \left (  1 - {{r^2}\over{b^2}} - {{N_e^2}\over{r^2}}\right )^{-1} dr^2
\pm\ 2N_e \sin\theta\ d\phi dt\ +\ {{r^4}\over{b^2}} d\Omega^2\ ,\label{dS4D-41}
\ee
\be
ds^2_{\rm QdT}=-\left ( 1- {{r^2}\over{b^2}} + {{N_e^2}\over{r^2}}\right ) dt^2 + \left (  1 - {{r^2}\over{b^2}} + {{N_e^2}\over{r^2}}\right )^{-1} dr^2
\pm\ 2N_e \sin\theta\ d\phi dt\ +\ {{r^4}\over{b^2}} d\Omega^2. \label{dS4D-41b}
\ee
In addition to a cosmological horizon, the Schwarzschild dS quantum black hole in eq(\ref{dS4D-41}) is characterized by an event horizon at
$r\rightarrow r_e$. A real spin angular momentum turns out to be sourced by an electric charge in the de Sitter black holes. The angular velocities are computed at the event horizon and are given by
\bea
&&\Omega^{\phi}_{\rm QdS}= {{\pm b^2N_e}\over{(b^2-N_e^2)^2\sin\t}} \nonumber\\
{\rm and}&&\Omega^{\phi}_{\rm QdT}={{\pm b^2N_e}\over{(b^2 +N_e^2)^2\sin\t}}\ .\label{dS4D-5a}
\eea
It ensures a higher spin for a quantum Schwarzschild de Sitter than a topological de Sitter black hole. A priori the angular velocities appear to vary with a polar angle. It is infinite at poles and takes a minimum value at an equator. The quantum de Sitter black holes presumably enforce a slicing in the polar angle. Nevertheless a need for the $\t$-slicing geometry may be resolved in presence of an auxiliary parameter $p$. Then, the angular velocities may be re-expressed as:
\bea
&&\Omega^{\phi}_{\rm QdS}= {{bp}\over{(b^2-N_e^2)^2}} \nonumber\\
{\rm and}&&\Omega^{\phi}_{\rm QdT}={b{p}\over{(b^2 +N_e^2)^2}}\ .\label{dS4D-5b}
\eea
For $Q_m=0$ the generic quantum black holes (\ref{dS4D-30}) and (\ref{dS4D-31}) may be simplified for its effective radial coordinate: 
$r_{\rm eff}=({\sqrt{r^4+p^2 \sin\t}})/b$. For a fixed $r$, it varies from a minimum $r^2/b$ to a maximum $({\sqrt{r^4+p^2}})/b$. 
In a limit $r_{\rm eff}\rightarrow 0$, the de Sitter quantum black holes shrink to a point singularity. Hence a curvature singularity at $r\rightarrow 0$ is removed by a two form fluctuation ($p\neq 0$) in the de Sitter stringy black holes. Similarly for $Q_m\neq 0$ and $p=0$ a Schwarzschild and a topological de Sitter stringy black holes are, respectively, described by a constant sphere of radii $({\sqrt{r^4-Q_m^2}})/b$ and 
$({\sqrt{r^4-Q_e^2}})/b$. The curvature singularity at $r\rightarrow 0$, in a Schwarzschild de Sitter quantum black hole and in a topological, is removed respectively by $r\rightarrow {\sqrt{Q_m}}$ and $r\rightarrow {\sqrt{Q_e}}$. 

%%%%%%%%%%%%%%%%%%%%%%%%%%%%%%%%%%%%%%%%%%%%%%%%%%%%%%%%%%%%%%%%%%%%%%%%%%%%%%
\subsection{Near horizon geometries: Limit ${\mathbf{\lambda \rightarrow 1}}$}
%%%%%%%%%%%%%%%%%%%%%%%%%%%%%%%%%%%%%%%%%%%%%%%%%%%%%%%%%%%%%%%%%%%%%%%%%%%%%% 
The effective de Sitter geometries, underlying a pair of $(D{\bar D})_3$-brane, may be analyzed in a limit $\lambda \rightarrow 1$ to address the origin of brane universe with a Big Bang. The limit implies $Q_e\rightarrow 0$. It does not a priori affect $Q_m$ and $p$ associated with the angular coordinates. However a change in signature underlying the Weyl scaling would imply $Q_m=0$ in the de Sitter vacuum at the Big Bang. In addition, the angular velocity becomes unphysical with an euclidean signature. It decouples from the de Sitter in the limit $r\rightarrow b$. In a 
limit $\lambda\rightarrow 1$ the de Sitter near cosmological horizon geometry (\ref{dS4D-5}) may be identified with the near horizon geometry of a
quantum black hole (\ref{bh-1}) on a pair of brane and anti-brane. In a global scenario the de Sitter vacua immediately after the Big Bang may be approximated for $N_e=0$ and $N_m\neq 0$. They reduce to yield:
\be
ds^2=\left ( 1- {{r^2}\over{b^2}}\right ) dt^2\ +\ \left (  1 - {{r^2}\over{b^2}}\right )^{-1}dr^2\ +\ \rho_m^2 d\Omega^2\ ,\label{BigBang-1}
\ee
where the effective radius $\rho_m=({\sqrt{r^4+Q_m^2}})/b$. A magnetic charge increases the radius of $S_2$ in a pure de Sitter vacuum. 
The de Sitter metric determinant is given by
\be
G= {{r^8}\over{b^4}}\left (1+{{Q_m^2}\over{r^4}}\right )^2 \sin^2\t = \rho_m^4\sin^2\t\ .\label{BigBang-2}
\ee 
The metric and the curvature are non singular in a limit $r\rightarrow b$. However for $Q_m=0$, the de Sitter vacua possess a curvature singularity in the limit of a cosmological horizon and is interpreted as a Big Bang singularity there. On the other hand the black hole (\ref{bh-1}) underlying a pre-Weyl scaling under an identical parameterization ($Q_e=0$ and $p=0$) may be given by
\be
ds^2=-\ \left (1 - {{M^2}\over{r^2}}\right ) dt^2 + \left ( 1 - {{M^2}\over{r^2}}\right )^{-1} dr^2\ +\ {{M^2}\over{r^2}}\rho_m^2 d\Omega^2\ .\label{BigBang-3}
\ee 
The black hole metric determinant in the case becomes
\bea
{\tilde G}&=&-\left (1+{{Q_m^2}\over{r^4}}\right )r^4 \sin^2\t\nonumber\\
&=&-{{b^4}\over{r^4}}\rho_m^4\sin^2\t\nonumber\\
&=&-\lambda^2 G\ .\label{BigBang-4}
\eea 
Formally a quantum black hole in lorentzian signature and a de Sitter vacuum in euclidean signature may seen to differ by a conformal factor $\lambda(r)$. The black hole is characterized by an event horizon at $r\rightarrow r_e=M$. Similar to de Sitter vacuum, the black hole metric and curvature are non-singular in the limit. It possesses a curvature singularity at $r\rightarrow 0$. However a curvature singularity is ruled out for a definite conformal scale $\lambda(r)$. In a limit $\lambda\rightarrow 1$, the difference in energy scale disappears. In the limit a black hole and a de Sitter respectively approach an event horizon and a cosmological horizon. Formally a coincident horizon at $r_c=r_e$, underlying two distinct vacua, may provoke thought to believe that the future time is presumably on a large $S^1$. In other words, a de Sitter in a non-perturbative formulation of quantum gravity may alternately be viewed through an exact Schwarzschild underlying a macroscopic black hole in Einstein gravity. Thus a thermal de Sitter (a small time) and a large time black hole may be related by a $T$-duality underlying $R\rightarrow {\alpha'}/R$ in string theory. It is further supported by a fact that a $4D$ thermal de Sitter quantum vacuum is conformally identified with a lorentzian Schwarzschild black hole sourced by a $5D$ metric potential in the formalism.

%%%%%%%%%%%%%%%%%%%%%%%%%%%%%%%%%%%%%%%%%%%%%%%
\section{Torsion geometries in ${\mathbf{4D}}$}
%%%%%%%%%%%%%%%%%%%%%%%%%%%%%%%%%%%%%%%%%%%%%%%
In this section we focus on a generalized brane geometry underlying a dynamical geometric torsion in the formalism. We obtain torsion geometries leading to a generalized $4D$ Reissner-Nordstrom (RN-) black hole which is essentially sourced by the zero and non-zero modes of a two form. The torsion potential may seen to describe the high energy (stringy) modes and hence they may be ignored in a low energy limit. The generalized (quantum) black hole in the limit shall be shown to reduce to a RN-vacuum in Einstein gravity.

\sp
\noindent
In the case, we consider an ansatz:
\bea
&&B_{t\theta}= B_{r\theta} =\ b\nonumber\\
{\rm and}&&B_{t\phi} =\ -M^2\cos\theta\ .\label{new-1}
\eea
Then the non-trivial components of a geometric torsion are worked out to yield:
\bea
&&{\cal H}_{t\theta\phi}\rightarrow H_{t\theta\phi} = -M^2\sin\theta\nonumber\\
{\rm and}&&{\cal H}_{tr\phi} = {{M^2b}\over{r^2}} \sin\theta\ .\label{new-2}
\eea
A geometric torsion incorporates metric fluctuations into the flat vacuum on a brane. Then the emergent metric on a vacuum created pair of 
$(D{\bar D})_3$-brane is sourced by the global modes of a NS two form in addition to the local modes of a two form in a non-linear $U(1)$ gauge theory. It is given by
\be
{G}_{\mu\nu}=\left ( g_{\mu\nu}\ \pm\ B_{\mu\lambda}g^{\lambda\rho}B_{\rho\nu}\ +\ C \ {\bar{\cal H}}_{\mu\lambda\rho}g^{\lambda\alpha}g^{\rho\beta}{\bar {\cal H}}_{\alpha\beta\nu}\right )\ .\label{emetricH}
\ee
For $C=(\pm 1/2)$ the line element is given by 
\bea
ds^2 =&-&\left(1 \pm \frac{b^2}{r^2} \pm \frac{M^4}{r^4} \pm \frac{M^4b^2}{r^6}\right)dt^2 +\left(1 \mp \frac{b^2}{r^2} \mp \frac{M^4b^2}{r^6}\right) dr^2+ \left(1  \mp \frac{M^4}{r^4}\right)r^2d\theta^2\nonumber\\ 
&+& \left(1 \mp \frac{M^4}{r^4} \mp \frac{M^4b^2}{r^6}\right)r^2\sin^2\theta d\phi^2\ \mp\ \frac{2b^2}{r^2} dtdr\ \pm\ \frac{2M^4b}{r^4}dr d\theta 
\ .\label{new-3}
\eea
%%%%%%%%%%%%%%%%%%%%%%%%%%%%%%%%%%%%%%
\subsection{Stringy RN-like de Sitter}
%%%%%%%%%%%%%%%%%%%%%%%%%%%%%%%%%%%%%%
In this section, we analyze the emergent geometries in a brane window for $r>(b,M)$. They shall be seen to describe the quantum Reissner-Nordstrom and Schwarzschild black holes on a pair of $(D{\bar D})_3$-brane in presence of an extra dimension. The quantum patches may be re-arranged in the regime on a vacuum created $D_3$-brane with $r^4>>b^4$ and $r^8>> M^8$ to yield:
\bea
ds^2&=&-\ \left( 1 \pm \frac{b^2}{r^2} \mp \frac{M^4}{r^4} \pm \frac{M^4b^2}{r^6}\right )dt^2\  +\ \left(1 \pm \frac{b^2}{r^2} \mp \frac{M^4}{r^4} \pm \frac{M^4b^2}{r^6}\right)^{-1} dr^2\  +\  r^2 d\t^2\nonumber\\ 
&&+\ \left ( 1\mp\frac{M^4b^2}{r^6}\right )r^2\sin^2\theta \ d\phi^2\ \mp\ \frac{2b^2}{r^2} \left ( dt-\frac{M^4}{br^2}d\theta\right ) dr\ \mp\ \frac{M^4}{r^4} ds^2_{\rm flat}\ ,\label{new-4} 
\eea
where $dS^2_{\rm flat}= \Big (2dt^2 + dr^2 + r^2d\Omega^2\Big )$. In fact the non-trivial quantum geometries are essentially sourced by a higher form gauge field or a two form on a $D_4$-brane which in turn describes a vacuum created pair of $(D{\bar D})_3$-brane. It enforces one to consider a global scenario underlying a pair. The relevant brane geometries on a pair of $(D{\bar D})_3$-brane become
\bea
ds^2&=&-\ \left( 1 \pm \frac{b^2}{r^2} - \frac{M^4}{r^4} + \frac{M^4b^2}{r^6}\right )dt^2\  +\ \left(1 \pm \frac{b^2}{r^2} - \frac{M^4}{r^4} + \frac{M^4b^2}{r^6}\right)^{-1} dr^2\  +\  r^2 d\t^2\nonumber\\ 
&&+\ \left ( 1-\frac{M^4b^2}{r^6}\right )r^2\sin^2\theta \ d\phi^2\label{new-5} 
\eea
and
\bea
ds^2&=&-\ \left( 1 \pm \frac{b^2}{r^2} + \frac{M^4}{r^4} - \frac{M^4b^2}{r^6}\right )dt^2\  +\ \left(1 \pm \frac{b^2}{r^2} + \frac{M^4}{r^4} - \frac{M^4b^2}{r^6}\right)^{-1} dr^2\  +\  r^2 d\t^2\nonumber\\ 
&&+\ \left ( 1+\frac{M^4b^2}{r^6}\right )r^2\sin^2\theta \ d\phi^2\ .\label{new-5b} 
\eea 
The first deformed term, in the causal patches, is sourced by the zero-modes in $B_2$-flux. It describes a Schwarzschild metric potential in $4D$ with an extra dimension. The second and third term correspond to the charges respectively underlying an axionic flux and a geometric torison in the formalism. In addition, the spherical symmetry is broken by a geometric torsion in the emergent quantum vacua. It may be recalled that a geometric torsion is primarily sourced by all the higher order terms underlying a $B_2$-coupling to an axion in a gauge theory. 

%%%%%%%%%%%%%%%%%%%%%%%%%%%%%%%%%%%%%%%%%%%%%%
\subsection{Non-propagating torsion geometries}
%%%%%%%%%%%%%%%%%%%%%%%%%%%%%%%%%%%%%%%%%%%%%%
In absence of a geometric torsion, one of the quantum geometries (\ref{new-5b}) precisely describes a quantum Reissner-Nordstrom black hole in $4D$ with a hint for a fifth dimension. Explicitly, the relevant quantum black hole is given by
\be
ds^2_{\rm QRN}=-\ \left( 1 - \frac{b^2}{r^2} + \frac{M^4}{r^4}\right )dt^2\  +\ \left(1 - \frac{b^2}{r^2} + \frac{M^4}{r^4}\right)^{-1} dr^2\  +\  r^2 d\Omega^2\ .\label{new-6} 
\ee
An extra dimension is intrinsic to a $B_2$-flux. It reassures a precise electro-magnetic charge, via a Poincare duality from an axionic charge $M$, in the Reissner-Nordstrom black hole. It shows that a constant two form components source a mass term and a local two form sources an electric (magnetic) charge. In a low energy limit, the quantum vacuum reduces to a typical Reissner-Nordstrom black hole in Einstein gravity. A formal identification of a quantum Reissner-Nordstrom vacuum with its (semi) classical geometry ensures an exact black hole in the formalism. 

\sp
\noindent
A priori three among the four quantum geometries (\ref{new-5})-(\ref{new-5b}), in absence of a geometric torsion, may seen to reduce to  
some of the known black holes in Einstein vacuum. The fourth geometry, under a flip of light cone at the event horizon, will be seen to identify 
with one among the remaining three in a perturbative gauge theory. Firstly, we focus on two of the relevant vacua and they are given by 
\be
ds^2=-\ \left( 1 - \frac{b^2}{r^2} \mp \frac{M^4}{r^4}\right )dt^2\  +\ \left(1 - \frac{b^2}{r^2} \mp \frac{M^4}{r^4}\right)^{-1} dr^2\  +\  r^2 d\Omega^2\ .\label{new-6a} 
\ee
They may be worked out for their renormalized black hole masses $b_{\rm ren}$ which formally leads to a 4D quantum Schwarzschild black hole in presence of an extra dimension in the formalism. It is given by
\bea
ds^2&=&-\ \left( 1 - \frac{b^2_{\rm ren}}{r^2}\right )dt^2\  +\ \left(1 - \frac{b^2_{\rm ren}}{r^2}\right)^{-1} dr^2\  +\  r^2 d\Omega^2\ ,\nonumber\\
{\rm where}\qquad b^2_{\rm ren}&=&b^2\left(1+\frac{N^2}{r^2}\right )_{r\rightarrow r_h=b_{\rm ren}}\quad {\rm and}\qquad N={{M^2}\over{b}}\ , \qquad\qquad {}\label{new-7}
\eea
The local degrees in a higher form gauge field is absorbed by the black hole mass to describe a dynamical vacuum. The computation of renormalized mass,
at the event horizon of each black hole, takes three different values ($b_s$, $b_+$ and $b_-$). Formally, they are:
\be
b_{\rm ren}={{b}\over{\sqrt{2}}}\left(1\pm {\sqrt{1 + {{4M^4}\over{b^4}}}}\right )^{1/2}\ .\label{new-7a}
\ee
Explicitly a renormalized mass $b_1$ of the Schwarzschild black hole (\ref{new-6a}) defined with $-M^4$ in $G_{tt}$ component becomes
\be
b_s={{b}\over{\sqrt{2}}}\left(1+ {\sqrt{1 + {{4M^4}\over{b^4}}}}\right )^{1/2}\ .\label{new-7b}
\ee
Interestingly the remaining geometry corresponds to a Reissner-Nordstrom black hole (\ref{new-6}). It may be re-expressed with two values of renormalized masses and they are:
\be
b_+={{b}\over{\sqrt{2}}}\left(1+{\sqrt{1 - {{4M^4}\over{b^4}}}}\right )^{1/2}\qquad 
{\rm and}\qquad b_-={{b}\over{\sqrt{2}}}\left(1-{\sqrt{1 - {{4M^4}\over{b^4}}}}\right )^{1/2}\ .\label{new-7c}
\ee
It re-assures the black hole horizons at $r\rightarrow r_{\pm}=b_{\pm}$. In other words, a Reissner-Nordstrom black hole may be interpreted as a multi-black hole solution in the formalism. The relative strength of the parameters ($b,M$) may further be explored to obtain simplified expression for the renormalized masses. Within a brane window, $i.e.\ r>(b,M)$ with $r^4>>b^4$ and $r^8>>M^8$, firstly we consider $b=M$. In the case, the geometry reduces to a background Schwarzschild black hole defined with an event horizon $b_{\rm ren}=b_s\rightarrow b$. The local degree underlying a torsion is dropped out in the regime. Secondly for $M>b$ and $M^2>>b^2$, the only sensible renormalized mass may be approximated by $b_s=M$ and describes a macroscopic black hole. Thirdly for $M<b$ and $M^8<<b^8$, the renormalized masses are approximated by $b_s=\Big (b + M^4/2b^3\Big )$, $b_+= \Big (b- M^4/2b^3\Big )$ and $b_-= \Big (M^2/b\Big )$ and they correspond to a macroscopic Reissner-Nordstrom black hole. 

\sp
\noindent
On the other hand a gauge theoretic torsion or its parameter $M$ underlying an axionic charge in $4D$ may alternately be identified with an electro-magnetic charge in presence of a fifth dimension. A renormalized non-linear charge $M_{\rm ren}$ is worked out by absorbing a cloud of zero modes of $B_2$-flux in its non-zero modes. In absence of a geometric torsion the brane geometry (\ref{new-5}) defined with $+b^2/r^2$ in $G_{tt}$ component may be re-expressed as:
\bea
ds^2&=&-\ \left( 1 - \frac{M^2_{\rm ren}}{r^4}\right )dt^2\  +\ \left(1 - \frac{M^2_{\rm ren}}{r^4}\right)^{-1} dr^2\  +\  r^2 d\Omega^2\nonumber\\
{\rm where}\qquad\quad M_{\rm ren}&=&M\left ( 1- {{r^2}\over{N^2}}\right )^{1/4}_{r\rightarrow r_h=M_{\rm ren}}\ .\label{new-8}
\eea
The renormalized mass of a Schwarzschild black hole mass (\ref{new-8}) may explicitly be worked out to yield:
\be
M_{\rm ren}= {{b}\over{\sqrt{2}}}\left ( -1 + \sqrt{1+ {{4M^4}\over{b^4}}}\right )^{1/2}\ .\label{new-8a}
\ee
Similarly in absence of a geometric torsion the brane geometry (\ref{new-5b}) defined with $+b^2/r^2$ in $G_{tt}$ component may be identified with a Schwarzschild black hole (\ref{new-7}) for its mass (\ref{new-7b}) under a flip of light cone at the event horizon.
The renormalized black hole mass may be approximated to yield $M_{\rm ren}=M$ and $M_{\rm ren}=\Big (M^2/b\Big )$, respectively, for $M>b$ and ($M<b$ with $M^8<<b^8$). Eqs.(\ref{new-7a}), (\ref{new-7b}) and (\ref{new-8a}), further imply $M_{\rm ren}<b_{\rm ren}$. Thus, the renormalized masses leading to the Schwazschild geometries (\ref{new-8}) respectively with mass $M$ and $\Big (M^2/b\Big )$ correspond to a microscopic black hole. It implies that a two form underlying its (renormalized) zero mode may define the mass of a classical Schwarzschild black hole in a non-perturbation geometric torsion formalism. In a low energy limit, all the local degrees in a geometric torsion are absorbed in a renormalized mass $b_{\rm ren}$ to define Einstein vacuum without any propagating torsion. A renormalized non-zero mode, of two form, defines a non-linear $U(1)$ charge in a quantum Schwarzschild black hole underlying a perturbation theory.

\sp
\noindent
Interestingly, a renormalized charge is worked out at an event horizon to obtain $M_{\rm ren}=b_{\rm ren}$. It establishes a fact that a Schwarzschild black hole is indeed unique. In fact, a precise identification of a renormalized charge with a (renormalized) mass of a black hole is an artifact of higher form $U(1)$ gauge theory on a $D_3$-brane. Thus, a Schwarzschild black hole in the formalism is precisely described in a perturbation gauge theory on a $D_3$-brane in absence of a geometric torsion. In other words the low energy limit reassures a non-propagating torsion in the formalism. As result the generalized curvature reduces to Riemannian geometry in the limit and is described by a perturbation theory. Then, the black hole mass in eq(\ref{new-7}) reassures a large `$M$' and a small `$b$' for a typical Schwarzschild black hole. In fact a large and a small value of parameter are indeed enforced by an underlying $U(1)$ gauge theory in $H_3$ defined with a small $B_2$-coupling (\ref{gauge-5}). A priori, the emergent Schwarzschild black hole (\ref{new-8}) defined with $M_{\rm ren}$ hints at three extra dimensions unlike to one extra dimension with $b_{\rm ren}$ in eq.(\ref{new-7}). However analysis, underlying a zero and a non-zero mode of a $B_2$-flux, reveals a fifth extra dimension. It resolves the apparent disparity in the number of extra dimensions between two alternate descriptions.

%%%%%%%%%%%%%%%%%%%%%%%%%%%%%%%%%%%%%%%%%%%%%%%%
\subsection{Stringy de Sitter black holes}
%%%%%%%%%%%%%%%%%%%%%%%%%%%%%%%%%%%%%%%%%%%%%%%%
Now we Weyl scale the emergent quantum geometries by  
$\lambda=(b^2/r^2)$ on a $D_3$-brane. Under $\lambda$ scaling, the quantum geometric patches sourced by a geometric torsion in the formalism may be described by
\bea
ds^2&=&\left( 1 -\frac{r^2}{b^2}  \pm \frac{N^2}{r^2} \mp \frac{M^4}{r^4}\right)dt^2\  +\ \left(1 + \frac{r^2}{b^2} \pm \frac{N^2}{r^2} \mp \frac{M^4}{r^4}\right) dr^2\  +\ \frac{r^4}{b^2} d\Omega^2\nonumber\\ 
&&\mp\ \frac{N^2}{r^2} \left( 2dt^2 + dr^2 + r^2 d\Omega^2 \right)\ \mp\ \frac{M^4}{r^2}\sin^2\theta\ d\phi^2\ +\ 2dtdr\ \pm\ \frac{2M^4}{br^2}dr d\theta\label{new-8a} 
\eea
\bea
{\rm and}\quad ds^2&=&-\ \left(1 +\frac{r^2}{b^2}  \mp \frac{N^2}{r^2} \pm \frac{M^4}{r^4}\right)dt^2\  -\ \left(1 - \frac{r^2}{b^2} \mp \frac{N^2}{r^2} \pm \frac{M^4}{r^4}\right) dr^2\  +\ \frac{r^4}{b^2} d\Omega^2\nonumber\\ 
&&\mp\ \frac{N^2}{r^2} \left( 2dt^2 + dr^2 + r^2 d\Omega^2 \right)\ \mp\ \frac{M^4}{r^2}\sin^2\theta\ d\phi^2\ -\ 2dtdr\ \pm\ \frac{M^4}{br^2}dr d\theta\ .\label{new-8b} 
\eea
We investigate a quantum regime for $M<r<b$ on a $D_3$-brane. Then, the metric component $G_{rr}$ may be approximated with $M^8<<r^8$ and $r^4<<b^4$. They may a priori be described by some of the AdS and de Sitter causal patches on a $D_3$-brane. However the causal patches, sourced by the local modes of a higher form, appears to be associated with a relative wrong sign. They are given by 
\bea
ds^2&=&\left( 1 -\frac{r^2}{b^2}  \pm \frac{N^2}{r^2} \mp \frac{M^4}{r^4}\right)dt^2\  +\ \left(1 - \frac{r^2}{b^2} \mp \frac{N^2}{r^2} \pm \frac{M^4}{r^4}\right)^{-1} dr^2\  +\ \frac{r^4}{b^2} d\Omega^2\nonumber\\ 
&&\mp\ \frac{N^2}{r^2} \left( 2dt^2 + dr^2 + r^2 d\Omega^2 \right)\ \mp\ \frac{M^4}{r^2}\sin^2\theta\ d\phi^2\ +\ 2dtdr\ \pm\ \frac{2M^4}{br^2}dr d\theta\label{new-9a} 
\eea
\bea
{\rm and}\quad ds^2&=&-\ \left(1 +\frac{r^2}{b^2}  \mp \frac{N^2}{r^2} \pm \frac{M^4}{r^4}\right)dt^2\  -\ \left(1 + \frac{r^2}{b^2} \pm \frac{N^2}{r^2} \mp \frac{M^4}{r^4}\right)^{-1} dr^2\  +\ \frac{r^4}{b^2} d\Omega^2\nonumber\\ 
&&\mp\ \frac{N^2}{r^2} \left( 2dt^2 + dr^2 + r^2 d\Omega^2 \right)\ \mp\ \frac{M^4}{r^2}\sin^2\theta\ d\phi^2\ -\ 2dtdr\ \pm\ \frac{2M^4}{br^2}dr d\theta\ .\label{new-9b} 
\eea
Under an interchange $dt_e\leftrightarrow dr$, the naive AdS causal patches (\ref{new-9b}) may be re-arranged to describe the de Sitter vacua (\ref{new-9a}). In a global scenario, $i.e.$ on a pair of vacuum created $(D{\bar D})_3$-brane, the quantum geometries may be re-expressed as:
\bea
ds^2&=&\left( 1 -\frac{r^2}{b^2}  \mp \frac{N^2}{r^2} \pm \frac{M^4}{r^4}\right)dt^2\  +\ \left(1 - \frac{r^2}{b^2} \pm \frac{N^2}{r^2} \mp \frac{M^4}{r^4}\right)^{-1} dr^2\  +\ \frac{r^4}{b^2} d\Omega^2\nonumber\\ 
&&\mp\ \frac{M^4}{r^2}\sin^2\theta\ d\phi^2 \mp \frac{N^2}{r^2} ds^2_{\rm flat}\ .\label{new-10} 
\eea
On the other hand, the causal patches in the emergent quantum geometries may be viewed through a $(2\times 2$) matrix ${\cal M}$ projection \cite{spsk-JHEP} on the column vectors: $\pmatrix{ {1} \cr {0}}$ and $\pmatrix{ {0} \cr {1}}$. The causal patches define the matrix. It is given by
\begin{equation}
{\cal M}=\frac{1}{2}\left( \begin{array}{ccc}
{{\tilde{G}}'}_{tt}(+)& & {{\tilde{G}}'}_{rr}(+)\\
 & & \\
{{\tilde{G}}'}_{rr}(-)& & {{\tilde{G}}'}_{tt}(-)
\end{array} \right)\ , \label{new-11}
\end{equation}
\bea
{\rm where}\quad {{\tilde G}'}_{tt}(+)&=& \left (1-{{r^2}\over{b^2}}+ {{N^2}\over{r^2}} -{{M^4}\over{r^4}}\right )\ ,\qquad\quad{{\tilde G}'}_{rr}(+)= \left (1-{{r^2}\over{b^2}} + {{N^2}\over{r^2}} - {{M^4}\over{r^4}}\right )^{-1}\ ,\qquad\qquad {}\nonumber\\ {{\tilde G}'}_{tt}(-)&=& \left (1-{{r^2}\over{b^2}} -  {{N^2}\over{r^2}}+{{M^4}\over{r^4}}\right )\quad {\rm and}\quad\;{{\tilde G}'}_{rr}(-)= \left (1-{{r^2}\over{b^2}} -{{N^2}\over{r^2}} + {{M^4}\over{r^4}} \right )^{-1}\ .\label{new-12}
\eea
The inverse of matrix becomes
\be
{\cal M}^{-1}=\frac{1}{2\det {\cal M}}\left( \begin{array}{ccc}
{{\tilde{G}}'}_{tt}(-) & & -{{\tilde{G}}'}_{rr}(+)\\
 && \\
-{{\tilde{G}}'}_{rr}(-)& &{{\tilde{G}}'}_{tt}(+)
\end{array} \right )\ .\label{new-13}
\ee
It is important to note that the matrix projection on two independent column vectors project the quantum causal patches obtained on a brane 
(\ref{new-10}). Explicitly, they are:
\be
{\cal M}\left( \begin{array}{c}
1\\
\\0
\end{array}\right)=\frac{1}{2}\left( \begin{array}{c}
{{\tilde{G}}'}_{tt}(+)\\
\\
{{\tilde{G}}'}_{rr}(-)
\end{array}\right) 
\qquad{\rm and} \qquad {\cal M}\left( \begin{array}{c}
0\\
\\
1
\end{array}\right)=\frac{1}{2}\left( \begin{array}{c}
{{\tilde{G}}'}_{rr}(+)\\
\\
{{\tilde{G}}'}_{tt}(-)
\end{array}\right) \ . \label{new-14}
\ee
Explicitly, the determinant of the matrix is worked out in the brane regime to yield:
\be
\det {\cal M}= -{{r^2}\over{b^2}}\ .\label{new-15}
\ee
The determinant at the cosmological horizon ensures ($\det {\cal M}=-1$) a discrete transformation, underlying a non-degenerate matrix ${\cal M}$
for its projection on two dimensional column vectors. We perform an inverse matrix operation on the column vectors to separate out the mixed causal patches in a quantum regime. They are given by
\be
{\cal M}^{-1}\left( \begin{array}{c}
1\\
\\
0
\end{array}\right)=\frac{1}{2}\left( \begin{array}{c}
-{{\tilde{G}}'}_{tt}(-)\\
\\
{{\tilde{G}}'}_{rr}(-)
\end{array}\right) 
\qquad{\rm and} \qquad {\cal M}^{-1}\left( \begin{array}{c}
0\\
\\
1
\end{array}\right)=\frac{1}{2}\left( \begin{array}{c}
{{\tilde{G}}'}_{rr}(+)\\
\\
-{{\tilde{G}}'}_{tt}(+)
\end{array}\right)\ . \label{new-16}
\ee
Then the inverse matrix projected quantum causal patches on a brane may appropriately be re-expressed as:
\bea
ds^2&=&-\left( 1 -\frac{r^2}{b^2}-\frac{N^2}{r^2} +\frac{M^4}{r^4}\right)dt^2\  +\ \left(1 - \frac{r^2}{b^2} - \frac{N^2}{r^2} +\frac{M^4}{r^4}\right)^{-1} dr^2\  +\ \frac{r^4}{b^2} d\Omega^2\nonumber\\ 
&&-\ \frac{M^4}{r^2}\sin^2\theta\ d\phi^2 \ +\ \frac{N^2}{r^2} ds^2_{\rm flat}\ .\label{new-17a}
\eea
\bea
{\rm and}\quad ds^2&=&-\left( 1 -\frac{r^2}{b^2}+\frac{N^2}{r^2} -\frac{M^4}{r^4}\right)dt^2\  +\ \left(1 - \frac{r^2}{b^2} + \frac{N^2}{r^2} -\frac{M^4}{r^4}\right)^{-1} dr^2\  +\ \frac{r^4}{b^2} d\Omega^2\qquad\qquad {}\nonumber\\ 
&&+\ \frac{M^4}{r^2}\sin^2\theta\ d\phi^2\ -\ \frac{N^2}{r^2} ds^2_{\rm flat}\ .\label{new-17b}
\eea
A priori, the causal patches reveal a quantum Reissner-Nordstrom de Sitter (\ref{new-17a}) and a quantum Schwarzschild de Sitter (\ref{new-17b}) geometries. In the case a defined parameter $N$ may be interpreted as a mass of black hole and $M$ defines a two form non-linear charge. However 
two among the three parameters  $(b,N,M)$ are independent. Thus an apparent Reissner-Nordstrom de Sitter vacuum is indeed an artifact of naive parameterization. In absence of the local degrees in a torsion, $i.e.$ for $M=0$, the emergent geometries reduce to a spherically symmetric de Sitter vacuum in 4D. Ignoring the flat metric terms in the quantum geometries, the non-trivial de Sitter vacua (\ref{new-17a}) and (\ref{new-17b}) may be given by
\bea
ds^2&=&-\left( 1 -\frac{r^2}{b^2}\right )\left (1 + \frac{M^4}{r^4}\right)dt^2\  +\ \left(1 - \frac{r^2}{b^2}\right )^{-1}
\left (1+\frac{M^4}{r^4}\right)^{-1} dr^2\ +\ {{r^4}\over{b^2}}d\t^2\nonumber\\
&&\quad\qquad\qquad\qquad\qquad\qquad\qquad\qquad\qquad\qquad +\ \left (\frac{r^2}{b^2}-\frac{M^4}{r^4}\right ) r^2\sin^2\theta\ d\phi^2\ .\label{new-18a}
\eea
\bea
{\rm and}\quad ds^2&=&-\left( 1 -\frac{r^2}{b^2}\right )\left (1 - \frac{M^4}{r^4}\right)dt^2\  +\ \left(1 - \frac{r^2}{b^2}\right )^{-1}
\left (1-\frac{M^4}{r^4}\right)^{-1} dr^2\ +\  {{r^4}\over{b^2}}d\t^2\qquad\qquad {}\nonumber\\
&&\quad\qquad\qquad\qquad\qquad\qquad\qquad\qquad\qquad\qquad 
+\ \left (\frac{r^2}{b^2}+\frac{M^4}{r^4}\right ) r^2\sin^2\theta\ d\phi^2\ .\label{new-18b}
\eea
The de Sitter vacuum (\ref{new-18a}) is defined with a cosmological horizon at $r\rightarrow r_c=b$. A local torsion breaks the $S_2$-symmetry and deforms the causal patches without any significant change and hence may be identified with a topological de Sitter black hole in 4D. The other 
de Sitter vacuum (\ref{new-18b}) possesses an event horizon at $r\rightarrow r_e=M$ in addition to a cosmological horizon at $r_c=b$. It describes a Schwarzschild de Sitter black hole in 4D. Most importantly, a superposition of two independent geometries re-assures a pure de Sitter vacuum, which is
essentially sourced by the zero modes of a two form in the formalism. The cancellation of non-zero modes contribution or the 
local degrees of a geometric torsion in the overlapping geometry ensures a gauge theoretic vacuum on a $D_4$-brane. In other words, a topological de Sitter and a Schwarzschild de Sitter quantum black holes are created at a Big Bang singularity $r_c=b$, respectively on a $D_3$-brane and a ${\bar D}_3$-brane. The black hole geometries are indeed sourced by a geometric torsion in the formalism.

\sp
\noindent
The nontrivial section in the quantum de Sitter (\ref{new-17a}) on a brane may also be re-expressed in terms of renormalized mass $N_{\rm eff}$ or a charge $M_{\rm eff}$. It takes a form:
\be
ds^2=-\left( 1 -\frac{r^2}{b^2} +\frac{N^2_{\rm eff}}{r^2}\right)dt^2  + \left(1 - \frac{r^2}{b^2} + \frac{N^2_{\rm eff}}{r^2}\right)^{-1} dr^2  
+ {{r^4}\over{b^2}} d\t^2 +\left (\frac{r^2}{b^2} - \frac{M^4}{r^4}\right )r^2\sin^2\theta d\phi^2\label{new-19a}
\ee
or
\be
ds^2=-\left( 1 -\frac{r^2}{b^2}+\frac{M^4_{\rm eff}}{r^4}\right)dt^2 +\left(1 - \frac{r^2}{b^2} +\frac{M^4_{\rm eff}}{r^4}\right)^{-1} dr^2 
+\frac{r^4}{b^2} d\t^2 + \left ({{r^2}\over{b^2}}-\frac{M^4}{r^4}\right )r^2\sin^2\theta d\phi^2\ ,\label{new-20a}
\ee
\be
{\rm where}\quad N^2_{\rm eff}= N^2\left ({{b^2}\over{r^2_e}}-1\right )\qquad {\rm and}\quad M_{\rm eff}\approx M
\left (1-{{r^2_h}\over{4b^2}}\right )\ .\qquad\qquad\qquad\qquad\qquad\qquad\;\;\;\;\;\; {}\label{new20}
\ee
A priori, the causal patches in the quantum geometry (\ref{new-19a}) may be identified with a four dimensional TdS black hole in presence of an extra dimension in Einstein vacuum. However a propagating torsion (\ref{new-2}) in the formalism ensures a non-Riemannian geometry underlying a quantum TdS black hole. The Ricci scalar takes a form:
\be
R= {{F(M,b,\theta)}\over{b^2r^6(b^2M^4-r^6)}}\ .\label{new-201}
\ee
A priori a quantum TdS possesses curvature singularities at $r\rightarrow 0$ and at $r\rightarrow (bM^2)^{1/3}$. Nevertheless a non-Riemannian torsion geometry leading to an effective TdS quantum black hole is described by a modified curvature scalar ${\cal K}$. The  curvature scalar is computed in a TdS quantum black hole which is sourced by a geometric torsion. It implies:
\be
R\ \rightarrow\ {\cal K}= {{M^4}\over{2r^4}}\left (1 + {{b^2}\over{r^2}}\right )\ .\label{new-202}
\ee
Naively a TdS quantum black hole possesses a curvature singularity defined in a limit $r\rightarrow 0$. However on a brane window $M<r<b$, a geometric torsion forbids the possibility of a curvature singularity in a TdS quantum geometry on a brane. A geometric torsion, leading to a quantum black hole in string theory, presumably manifests a gravitational repulsion that prevents the formation of singularities in the vacuum created brane/anti-brane at a Big Bang. Interestingly, the emergent black hole on a brane is characterized by an event horizon at 
\be
r\rightarrow r_h \approx b\left ( 1 + {{N^2_{\rm eff}}\over{2b^2}}\right )\ .\label{new-203}
\ee 
In a near horizon limit $r\rightarrow b$, the curvature scalar takes a smallest positive constant value and further reconfirms a de Sitter geometry on a vacuum created pair of $(D{\bar D})_3$-brane.

\sp
\noindent
Similarly, ignoring a flat section, the de Sitter black hole (\ref{new-17b}) may be re-expressed in terms of its renormalized mass or its renormalized charge. It is given by
\be
ds^2=-\left( 1 -\frac{r^2}{b^2} -\frac{N^2_{\rm eff}}{r^2}\right)dt^2  + \left(1 - \frac{r^2}{b^2} - \frac{N^2_{\rm eff}}{r^2}\right)^{-1} dr^2  
+ {{r^4}\over{b^2}} d\t^2 +\left (\frac{r^2}{b^2} + \frac{M^4}{r^4}\right )r^2\sin^2\theta d\phi^2\label{new-19b}
\ee
or
\be
ds^2=-\left( 1 -\frac{r^2}{b^2}-\frac{M^4_{\rm eff}}{r^4}\right)dt^2 +\left(1 - \frac{r^2}{b^2} -\frac{M^4_{\rm eff}}{r^4}\right)^{-1} dr^2 
+\frac{r^4}{b^2} d\t^2 + \left ({{r^2}\over{b^2}}+\frac{M^4}{r^4}\right )r^2\sin^2\theta d\phi^2\ .\label{new-20b}
\ee
It describes a four dimensional SdS quantum geometry sourced by a geometric torsion on a vacuum created pair of $(D{\bar D})_3$-brane. Though the curvature singularity is absent in a brane regime, the geometry describes a black hole with a cosmological horizon $r_c$ and an event horizon $r_e$. Explicitly, they may be approximated to yield:
\be
r_c \approx b\left ( 1 - {{N^2_{\rm eff}}\over{2b^2}}\right )\qquad {\rm and}\quad r_e\approx N_{\rm eff}\ .\label{new-204}
\ee 
The horizon radii in the quantum black holes (\ref{new-20a}) and (\ref{new-20b}) may also be estimated by their renormalized charge $M_{\rm eff}$. The  emergent TdS and SdS geometries may respectively be viewed through their origin in a vacuum created $D_3$-brane and an anti $D_3$-brane or vice-versa at a cosmological horizon $r_c=b$, which was shown to describe a Big Bang singularity by the authors in ref.\cite{spsk-JHEP}. In fact, it was argued that the brane-Universe(s) began instantaneously underlying a vacuum created pair of D-instantons by a higher form gauge field on a $D_4$-brane.
The coupling of a dynamical torsion, $i.e.\ M\neq 0$, lead to two different quantum vacua underlying a TdS and a SdS black hole. Qualitatively, the event horizon in both cases may be described at $r\rightarrow {\sqrt{b}}$. However, a TdS black hole corresponds to an upper cut-off $b<1$ and a SdS black hole is defined with a lower cut-off $1<b$. In other words, a geometric torsion begins to interplay at $b=1$ and distinguishes a TdS quantum black hole from a SdS quantum black hole. The geometric torsion also breaks the spherical symmetry. The metric potential signifies an extra fifth dimension between a pair. Interestingly a geometric torsion sources a high energy mode and ensures an underlying stringy aspects in the emergent TdS and SdS geometries on a non-BPS brane.

\sp
\noindent
In a low energy limit, the torsion contribution becomes insignificant in the de Sitter black holes. In the limit, a TdS and a SdS are 
approximated by the Riemanian curvature and are respectively given by
\be 
ds^2=-\left( 1 -\frac{r^2}{b^2} -\frac{N^2}{r^2}\right)dt^2\  +\ \left(1 - \frac{r^2}{b^2} - \frac{N^2}{r^2}\right)^{-1} dr^2  
\ +\ {{r^4}\over{b^2}} d\Omega^2\label{new-21a}
\ee
\be
ds^2=-\left( 1 -\frac{r^2}{b^2} +\frac{N^2}{r^2}\right)dt^2\  +\ \left(1 - \frac{r^2}{b^2} + \frac{N^2}{r^2}\right)^{-1} dr^2\  
+\ {{r^4}\over{b^2}} d\Omega^2\ .\label{new-21b}
\ee
The spherical symmetry is restored in a SdS and a TdS black hole in absence of a high energy mode in the quantum geometries. The emergent geometries in a low energy limit correspond to classical vacua presumably in Einstein gravity. The correct curvature in the classical regime is described by a Ricci scalar. It is worked out for a SdS black hole (\ref{new-21a}) and is given by
\be
R={{20}\over{b^2}}\ +\ {{6N^2}\over{r^4}} \ +\ {{2(b^2-r^2)}\over{b^2r^4}}\left (b^2-7r^2\right )\ .\label{new-22}
\ee
The curvature singularity at $r\rightarrow 0$ is forbidden by the black hole mass $N$ in the brane window.  
A priori the cosmological horizon $r_c=b$ appears to shrink and an event horizon $r_e=N$ expands in a SdS black hole underlying a flow to set an equipotential sourced by a zero mode and a non-zero mode of two form. It leads to a Nariai black hole and observer lies along a time-like $r$.
A limit $r_e\rightarrow r_c$, along a spatial $r$, or generically $r\rightarrow r_c$ may be analyzed for the Ricci scalar in a Nariai black hole to yield:
$R=({26}/{b^2})$. Interestingly, the near horizon geometry in a TdS black hole (\ref{new-21b}) may be approximated by $R=({20}/{b^2})$. It is remarkable to note that a quantum TdS in a low energy limit reduces to a classical SdS presumably underlying Einstein vacuum. Similarly, a quantum SdS in the limit reduces to a classical TdS black hole on a vacuum created pair of $(D{\bar D})_3$-brane. The aspects of quantum tunneling of a SdS vacuum to a TdS, via a Nariai black hole, argued in ref.\cite{spsk-JHEP} may be invoked to resolve an apparent puzzle underlying a geometric transition between a quantum TdS (or SdS) and a classical SdS (or TdS).

%%%%%%%%%%%%%%%%%%%%%%%%%
\section{Concluding remarks}
%%%%%%%%%%%%%%%%%%%%%%%%%%%%
An effective geometric torsion dynamics, underlying an irreducible curvature scalar in a second order formalism, is explored on a $D_4$-brane. In particular, we have considered a two form in a $U(1)$ gauge theory on a $D_4$-brane in presence of a background metric which presumably describes an effective open string metric on an anti $D_4$-brane. In other words the background metric is assumed to be sourced by the global modes of a NS two form in an open string world-sheet. It was argued that a local two form can generate a pair of $(D{\bar D})_3$-brane at the cosmological horizon of a background de Sitter vacuum with a Big Bang. The non-linear charge, underlying a dynamical two form, is believed to generate an effective geometry on a brane and an anti-brane in a pair. However an observer in a brane universe is unaware of the existence of an anti-brane. Interestingly the effective curvature formalism on a $D_4$-brane turns out to be non-perturbative due to the $D$-instanton correction. Alternately a non-trivial geometry may be viewed by a pair of vacuum created $(D{\bar D})_3$-brane. Our analysis reveals an extra fifth dimension between a brane and an anti-brane worlds. The hidden dimension to a $D_3$-brane universe has been argued to play a significant role to explain the inflation observed in cosmology. Importantly the Poincare dual of a torsion, $i.e.$ an axion, on an anti $D_3$-brane has been argued to describe a quintessence in the formalism. Its dynamics on an anti $D_3$-brane greatly influences the effective $D_3$-brane universe through the hidden fifth dimension. A quintessence is believed to be a potential candidate for the dark energy. It accounts for a time varying vacuum energy density observed by CMB in cosmology. 

\sp
\noindent
The formalism is explicitly worked out to describe various effective de Sitter quantum geometries underlying Schwarzschild, topological and Reissner-Nordstrom black holes. The quantum black holes were shown to be free from curvature singularity at $r\rightarrow 0$ due to a lower bound on a brane window. In a low energy limit the nonperturbative correction sourced by a torison has been argued to be insignificant. The quantum black holes undergo expansion in the limit and reduce to Einstein vacuum. In the context a non-propagating two form on a $D_2$-brane may seen to possess its origin in a dynamical two form on a higher dimensional $D_p$-brane. Thus the existence of BTZ black hole in three dimensions due to a non-zero (negative) vacuum energy density in Einstein gravity may be addressed by a dynamical two form. It may lead to an effective curvature on a $D_p$-brane for $p\ge 3$ underlying a ten dimensional type IIA or type IIB superstring theory on $S^1$. In principle the formalism may be generalized to investigate various  higher form dynamics on a space filling $D_9$-brane in superstring theory. It would like to hint for the existence of an anti $D_9$-brane. Arguably the global geometries on a vacuum created pair of $(D{\bar D}_9$-brane may provide clue to conjectured M-theory in eleven dimensions.  

%%%%%%%%%%%%%%%%%%%%%%%%%%
\section*{Acknowledgments}
%%%%%%%%%%%%%%%%%%%%%%%%%%
Authors would like to thank Daya Shankar Kulshreshtha, Sudhakar Panda, Swarnendu Sarkar and Ashoke Sen for various useful discussions. R.K. gratefully acknowledges the CSIR fellowship.

%**********************************************************%
\def\anp{Ann. of Phys.}
\def\cmp{Comm.Math.Phys} {}
\def\prl{Phys.Rev.Lett.}
\def\prd#1{{Phys.Rev.}{\bf D#1}}
\def\jhep{JHEP\ {}}{}
\def\cqg#1{{Class.\& Quant.Grav.}}
\def\plb#1{{Phys. Lett.} {\bf B#1}}
\def\npb#1{{Nucl. Phys.} {\bf B#1}}
\def\mpl#1{{Mod. Phys. Lett} {\bf A#1}}
\def\ijmpa#1{{Int.J.Mod.Phys.}{\bf A#1}}
\def\mpla#1{{Mod.Phys.Lett.}{\bf A#1}}
\def\rmp#1{{Rev. Mod. Phys.} {\bf 68#1}}
\def\jaat{J.Astrophys.Aerosp.Technol.\ {}} {}

%**********************************************************%

\end{document}